\documentclass[]{aastex63}

\usepackage{comment}

\usepackage{graphicx}	
\usepackage{amsmath}	
\usepackage{amssymb}	
\usepackage{subfigure}

\accepted{for publication in AJ on April 9th, 2021}

\shorttitle{Eccentricity-color correlation}
\shortauthors{Ali-Dib et al.}

\graphicspath{{./}{figures/}}

\begin{document}

\title{The rarity of very red TNOs in the scattered disk}

\correspondingauthor{Mohamad Ali-Dib}
\email{mma9132@nyu.edu}

\author[0000-0002-6633-376X]{Mohamad Ali-Dib}
\affiliation{Institut de recherche sur les exoplan\`etes, Universit\'e de Montr\'eal, 2900 boul. \'Edouard-Montpetit, Montr\'eal, H3T 1J4, Canada}
\affiliation{Center for Astro, Particle and Planetary Physics (CAP3),
New York University Abu Dhabi, UAE}
\affiliation{Canadian Institute for Theoretical Astrophysics, 60 St. George St., Toronto, ON M5S 3H8, Canada}

\author[0000-0001-8617-2425]{Michaël Marsset}
\affiliation{Department of Earth, Atmospheric and Planetary Sciences, MIT, 77 Massachusetts Avenue, Cambridge, MA 02139, USA}

\author{Wing-Cheung Wong}
\affiliation{Centre for Planetary Sciences, University of Toronto Scarborough, Toronto, Ontario M1C 1A4, Canada}

\author{Rola Dbouk}
\affiliation{Department of Physics, American University of Beirut,
PO Box 11-0236, Riad El-Solh, Beirut 11097 2020, Lebanon}


\begin{abstract}

We investigate the origins of the photometrically Very Red and Less Red Trans-Neptunian Objects. We first reanalyse the dataset of \cite{marsset} and find that, in addition to the known color-inclination correlation in hot TNOs, a similar trend exists for color-eccentricity. We show that VR TNOs are  sharply constrained to eccentricities $<$ 0.42 and inclinations $<$ 21$^\circ${, leading to a paucity of VR scattered disk and distant MMR objects.} We then interpret these findings using N-body simulations accounting for Neptune's outward migration into a massless particles disk, and find that these observations are best reproduced with a LR-to-VR color transition line between $\sim$ 38 and 42 AU in the primordial disk, separating the objects' formation locations. {For an initial surface density profile ($\Sigma \propto 1/r^2$), a color transition around 38 AU is needed to explain the high abundance of VR plutinos but creates too many VR scattered disk objects, while a transition line around 42 AU seems to better reproduces the scattered disk colors but creates virtually no VR plutinos.} Our simulations furthermore show that the rarity of VR particles at high eccentricity is possibly due to the absence of sweeping higher order MMRs, and secular resonances, beyond 42 AU. Inspecting individual populations, we show that the majority of VR SDOs originate as objects trapped in Neptune's second and third order MMRs. These then evolve due to diffusion, scattering, Kozai-Lidov cycles, and secular resonances into their current orbits. Future unbiased color surveys are crucial to better constrain the TNOs dynamical origins.

\end{abstract}


\section{Introductions}
Trans-Neptunian Objects (TNOs) are fossils from the early days of the solar system, and give us a wealth of information about its formation and evolution. In particular, the orbits of these objects can be used to constrain the dynamical history of the giant planets, and their chemical composition can be linked to that of the protosolar nebula in which they formed. Observationally, multiple TNOs dynamical groups have been identified and characterized. These include first the classical Kuiper belt situated between 42 and 48 AU, with two subcomponents: the cold classicals with inclinations less than $\sim$ 5-6 degrees, and the hot classicals with higher inclinations and more complex dynamical histories. Second are the resonant TNOs, consisting of objects trapped in a mean motion resonance (MMR) with Neptune, with the 3:2 MMR (Plutinos) at 39.5 AU containing the largest number of known TNOs. The third major TNOs dynamical group is the scattered disk (SDOs), consisting of high eccentricity and semimajor axis objects scattered outward after close encounters with Neptune. These have mostly perihelions between 30 and 40 AU. Finally, the detached objects are extreme SDOs that are now largely detached from the gravitational influence of Neptune. We refer the reader to the reviews of \cite{glad2008,rev3,rev1,rev2} and the references therein for more detailed informations on the Kuiper belt and its structures.  

Many Solar System dynamical evolution models have been invoked to try to explain the eccentricity and inclination structure of TNOs over the years. While these models can now account for the overall shape of these structures, we are still far from fully understanding how did the Kuiper Belt evolve into its current form. The first dynamical group to unveil some of its secrets were the resonant objects, where \cite{malho1,malho2,malho3} showed these to be a natural consequence of Neptune's outward migration over multiple astronomical units to its current orbit. Neptune's migration also explains the origins of SDOs, as objects that were scattered outwards by Neptune during its migration instead of getting trapped in a resonance, colliding with a planet, or escaping the Kuiper Belt  \citep{duncanlev, luu}.  The cold classical belt on the other hand was originally thought to also be a result of Neptune's migration \citep{cold4}. Subsequent studies of the abundance and dynamical fragility of binaries in this population however strongly implied that the majority of cold classicals must have formed in-situ \citep{cold1,cold2,cold3,fraser2017,nesv2019c}. Finally, the formation of the hot classical population is yet to be fully understood, as models have not yet converged on an interpretation for its higher-end of the inclination distribution. While \cite{nesv2015} initially proposed that this might be due to a very slow migration of Neptune, \cite{volk} proposed that the giant planets secular architecture plays a more important role. {More recently, \cite{nesvadd1} and \cite{nesvadd2} pointed out that these differences can be explained by the different assumptions for Neptune's eccentricity evolution. While \cite{volk} adopted a very low eccentricity and considered higher inclinations migration of Neptune, \cite{nesv2015}'s results apply to a regime of migration where Neptune eccentricity is modestly excited to $\sim$ 0.1 and then damped. The $\nu8$ resonance acts in this case to implant bodies in the Kuiper belt.}

While simple models where Neptune migrates outward on a $\sim$ 10 Myr timescale into a mildly stirred up primordial planetesimals disk reproduces an overall sketch of the Kuiper Belt \citep{hahn2005}, these usually fail to explain the finer details of these structures. Smooth migration models for example lead to an overabundance of resonant particles, while in reality non-resonant orbits are more common by a factor 2-4 \citep{glad2012}. This led \cite{nesv2016} to propose that Neptune's migration was ``grainy'', physically caused by interacting with a smaller number of Pluto-size bodies, in contrast with smooth migration. \cite{kaib2016} then investigated the effects of such migration on high-perihelion objects near the 3:1 MMR. Another peculiar feature of the Kuiper belt is the ``kernel'', a concentration of dynamically cold orbits around a = 44 AU \citep{cold2}. While the origins and exact nature of this kernel is still far from being understood, \cite{nesv2015a} proposed that it might be due to a sudden jump in Neptune's semimajor axis during its migration due to a global instability in the solar system. More recently, \cite{gomes2020} proposed that the kernel could be caused by an initial sharp edge around 44.5 AU, and subsequent diffusion of cold classical bodies from inside that edge.

Another line of evidence that can shed light on the dynamical history of the Kuiper Belt and the giant planets is the chemical composition of TNOs, observed in proxy through their surface colors. 
One of the earlier such studies was \cite{color1,color2} who observed a ``gray-red'' (now termed ``Less Red'' (LR) and ``Very Red'' (VR)) color bimodality in a sample of $\sim$ 15 then 91 diverse KBOs. This was followed by \cite{peix2003} who showed a bimodality in the colors of centaurs in particular, where roughly half of their $\sim 20$ objects sample were blue–gray, and the rest were very red. Using a large survey of 109 objects, \cite{peix2004} noticed a correlation between the inclination and color of the classical belt objects. This was confirmed further in \cite{peix2008} who found that hot classicals with inclinations $\geq 10^\circ$ were mostly blue, while cold classicals are dominantly VR. The cold classicals were examined furthermore by the  Colours of the Outer Solar System Origins Survey (Col-OSSOS;  \citealt{schwamb2019}) who found them to be almost entirely VR \citep{peix2004,fraser2017}, with distinct surface characteristics from the much rarer dynamically excited VR objects \citep{pike2017}. The hot classicals were also reexamined through surveys by \cite{peix2012,fraser2012,wong2017} who concluded that these separately also show a strong color bimodality with the same shape as that found in centaurs. Finally, \cite{marsset} (hereafter M2019) also examined a sample of hot classicals, centaurs, resonant, and scattered objects using a dataset based on (but not exclusively) Col-OSSOS data for which discovery biases were modelled, and reported that VR TNOs are strongly limited to inclinations less than $\sim 21^\circ$. They hence concluded that a strong color-inclination correlation exists in this dataset, that persists even when considering the different sub-components individually.  They however stated that such correlation does not exist for eccentricity. 

In section \ref{dataansec} of this paper, we first reanalyse the dataset of M2019, and find an overlooked strong correlation between the eccentricity and color of hot classicals caused by different objects than those responsible for the color-inclination correlation. We then show that VR TNOs in this sample are strongly limited to eccentricities less than $\sim 0.42$. The scattered disk is hence found to be strongly dominated by LR objects, as hinted to by \cite{peix2004} and corroborated by M2019. This is in addition to the strong paucity of LR objects at inclinations $>$ 21 deg found by M2019. We conclude that VR TNOs are strongly contained to orbits with values below these two values in e-inc space. In section \ref{modelsec}, we model and interpret these observations using N-body simulations, and investigate the dynamical histories of the VR contaminants in the dominantly LR high eccentricity or inclination populations. We finally summarize our results and conclude in section \ref{conclusion}.

\section{A reanalysis of the dataset of Marsset et al.}
\label{dataansec}
\subsection{Eccentricity vs color}
\label{sec1}
\begin{figure*}
\begin{centering}
            \subfigure{\includegraphics[scale=0.38]{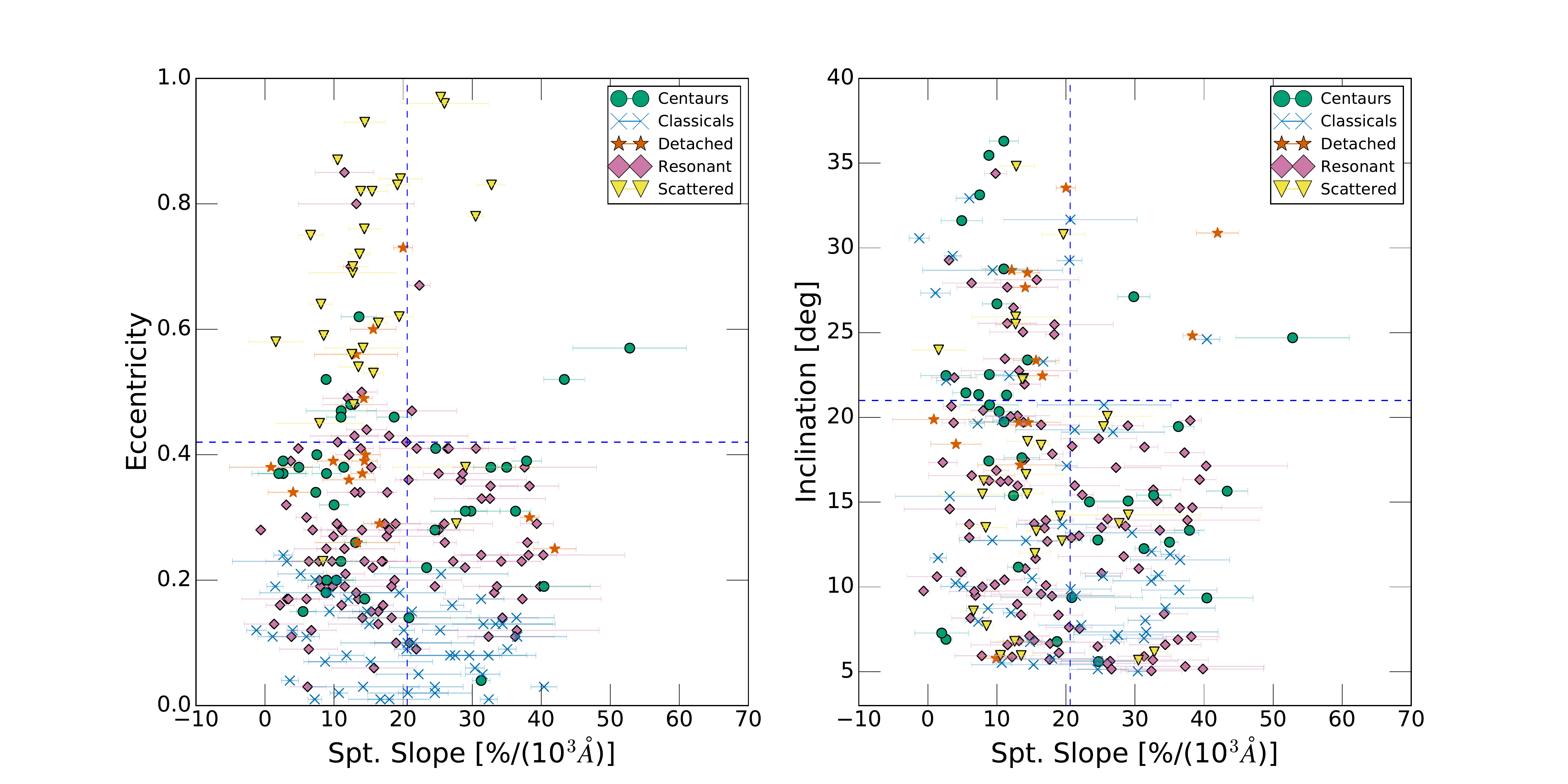}} 

   \caption{Orbital eccentricity and inclination vs spectral slope of the M2019 dataset, also showing the dynamical groups of the objects that we limit to 5 broad categories.
  LR (gray, neutral) objects are to the left of the vertical dashed line, with Spt. slope less than 20.6 [\%/(10$^3$Å)]) according to their photometric surface color, while VR (red) objects are to the right of the line. Both orbital quantities show a strong correlation with color, where VR objects are strongly limited to inclinations less than 21$^\circ$ as noticed by M2019, but also to eccentricities less than 0.42. The observed VR to LR ratio of low eccentricity or inclination objects is around unity. While the high inclination LR objects are a mix of different dynamical populations, the high eccentricity LR objects are dominated by the scattered disk and high order mean motion resonances.  }
    \label{fig:ecc1}
    \end{centering}
 \end{figure*}

In Fig. \ref{fig:ecc1} we plot the eccentricities and inclinations of the M2019 sample as a function of the spectral slopes. Both distributions show strong bimodality. First, as reported by M2019, TNOs with inclinations higher than $\sim$ 21$^{\circ}$ are almost exclusively LR (also referred to as gray or neutral objects), with an observed LR/VR ratio of $\sim$ 8.16 if all bodies are included. Second, we find that the eccentricity distribution clearly shows a similar behavior, where objects with eccentricities larger than $\sim$ 0.42 are predominantly LR with a LR to VR ratio of 5.37.  
This plot also shows that  a) the LR (aka ``gray'', ``neutral'') class is only found at e$>$0.4 for distant populations (scattered, detached, distant resonant), and b) the VR (aka ``red'') class is almost never found at e$>$0.4 for these distant populations. This is discussed in more details in the next subsections. 

We statistically test if the VR and LR populations have different eccentricity distributions using using Kolmogorov–Smirnov (KS) and Anderson–Darling (AD) tests. 

By considering the LR and VR sub-samples, a 2-sample KS test on the eccentricity distributions returns a statistic of 0.19. The two samples Anderson-Darling on the other hand returns a statistic of 4.19. 

To calculate the p-values of these statistics
we freeze the LR eccentricity sample, then repeatedly bootstrap with replacement a simulated VR sample from the frozen LR sample of equal size as the observed VR sample, then calculate the probability that a random sample produces a test statistic larger than the observed population. 
This provides a direct measure of the distribution of the statistic under the hypothesis that all objects are drawn from the LR sample.


We finally obtain p-values of 0.23\% for the KS test, and 0.085\% for the AD test, implying a $\geq$ 99\% probability that the LR and VR sub-samples have distinct eccentricity distributions. Both statistical tests therefore strongly suggest that the LR and VR sub-samples have two distinct distributions.

Now we redo the same analysis, but accounting for the error bars on the color measurements. Our method consist of bootstrapping the color and eccentricity distributions, but while replacing each color measurement with a value uniformly sampled from its entire uncertainty interval. New color values are sampled with each bootstrap sampling. We now find the KS and AD tests p-values to be 0.2\% and 0.05\%, respectively. The two color sub-populations hence are still very probably distinct even when accounting for the observational error bars. 

Finally, the same analysis can be redone while excluding centaurs from the dataset, as these are a transient population. For TNOs with inclinations higher than $\sim$ 21$^{\circ}$ we get a LR/VR ratio of 9.25, and for eccentricities larger than $\sim$ 0.42 we get a LR to VR ratio of 6.16.
For these cases, the color-eccentricity 2 samples KS and AD tests statistics are respectively 0.20 and 3.92, with p-values of 0.3\% and 0.05\%  when accounting for the error bars.



%

\begin{figure}
\centering
\includegraphics[scale=0.99]{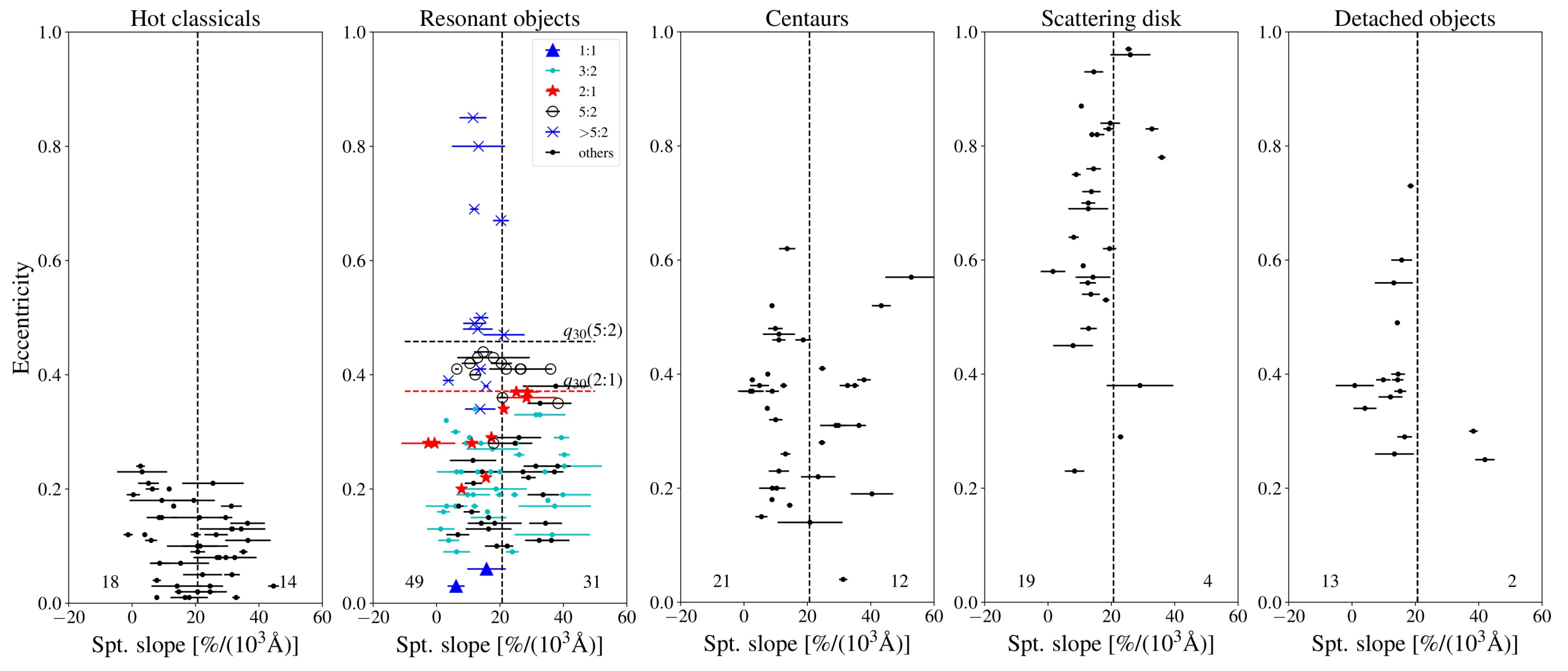}
   \caption{Orbital eccentricity vs spectral slope of the M2019 dataset, for 5 dynamical subpopulations. The color-eccentricity correlation is driven mainly by scattered disk objects, but also high order MMR objects {\color{black}where $q=30~au$ allows the objects to reach $ecc>0.42$.}}
\label{fig:eccMichael}
\end{figure}

\subsection{Eccentricity vs inclination}

\begin{figure*}
\begin{centering}

  {\includegraphics[scale=0.26]{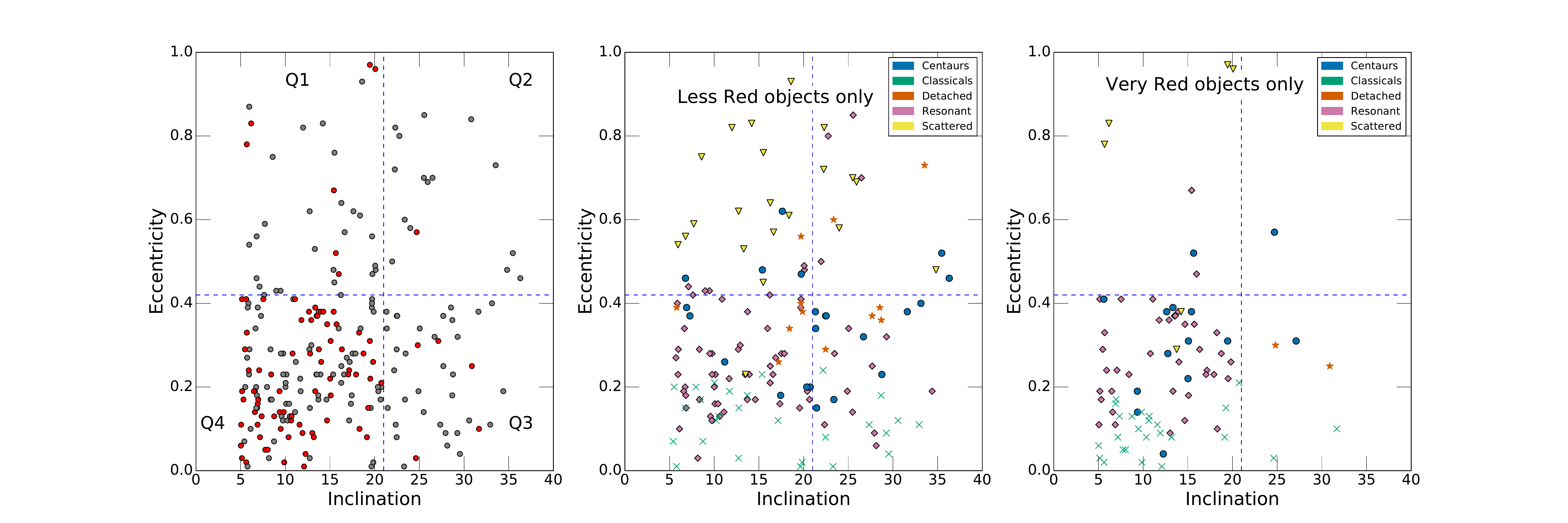}} 

        \end{centering}

   \caption{{Left panel:} The M2019 Col-OSSOS sample shown as an eccentricity-inclination plot. While the high inclination and high eccentricity populations share the objects of quadrant Q2, each has its own unique objects in quadrants Q3 and Q1 respectively. {Center and right panels:} Same as above, but showing the dynamical classifications of the objects, with the LR and VR objects plotted separately. Notice how VR objects are strongly constrained to Q4.}
    \label{fig:ecc3}
\end{figure*}

Now we investigate whether the eccentricity-color and inclination-color correlations are related, as would be the case for example if the eccentricities and inclinations of a given dominant population are highly correlated. In Fig. \ref{fig:ecc3} we plot the data on an eccentricity-inclination diagram, while showing the LR vs VR populations (left), and then separating them into different dynamical groups (center and right). We divide these plots into 4 quadrants. Q1 and Q2 contain objects with $e\geq0.42$, while Q2 and Q3 contain objects with $inc\geq 21^\circ$. The LR to VR ratios of these 4 quadrants are respectively 3.125, 16.0, 5.0, and 0.93. Hence the sole quadrant with near unity ratio is Q4 containing the dynamically colder objects.  We conclude that VR TNOs are strongly limited in eccentricity-inclination space to below e = 0.42 and inc = 21 deg.
The high eccentricity or inclination populations only have Q2 in common (objects with simultaneously high eccentricity and inclination), and hence it is worth investigating whether the color correlations survive if we exclude these objects.
First, by inspecting Q3 ($inc\geq 21^\circ$ but $e\leq 0.42$), it is clear that this population is diverse, with multiple dynamical groups contributing to the inclination-color correlation as found by M2019. Inspecting Q1 however ($inc\leq 21^\circ$ but $e\geq 0.42$), we notice that they are predominantly scattered disk objects (the roles of individual subpopulations are discussed in the next subsection). 

When excluding Q2, the eccentricities 2-samples (LR and VR) KS test statistic drops to 0.15 with a p-value of 2.35\%. It is hence statistically possible, but not certain, that the eccentricity-color correlation holds when excluding high inclination objects. If it does, this would imply that the two correlations have different physical origins, and are not due to eccentricity-inclination correlated sub-populations. Performing a Spearman correlation test on the eccentricties and inclinations of the entire dataset, we find a correlation coefficient of 0.18, with a p-value of 0.45\% . This hints that only a modest correlation exists between the two quantities in the dataset.

\subsection{Individual populations}

{ In Fig. \ref{fig:ecc1} we also show the different dynamical groups for all objects in the dataset. For ease of comparison, we additionally plot the eccentricity vs spectral slope of the M2019 dataset objects, for the 5 dynamically distinct subpopulations, in Fig. \ref{fig:eccMichael}.  To further test if the eccentricity-color correlation is dominated by a single dynamical subpopulation, we will repeat the statistical tests of section \ref{sec1} for each individual dynamical groups. The results are summarized in table \ref{indivpops}. We have excluded the classical Kuiper belt objects as all of them have eccentricities lower than 0.2, and are distributed almost evenly between LRs and VRs.

\begin{table*}[]

\begin{centering}

\begin{tabular}{|l|l|l|}
\hline
Population & KS statistic, p-value & AD statistic, p-value \\ \hline
All        & 0.19,   0.2\%         & +4.19,   0.05\%       \\ \hline
Centaurs   & 0.25,   18.6\%        & -0.69,   61.2\%       \\ \hline
Resonants  & 0.12,   46.5\%        & -0.29,   85.3\%       \\ \hline
SDOs       & 0.40,   17.5\%        & +1.18,   7.9\%        \\ \hline
\end{tabular}
\caption{The KS and AD tests statistics and p-values calculated for individual populations in the dataset of M2019.}
\label{indivpops}

\end{centering}
\end{table*}

\subsubsection{Centaurs}
We first start with centaurs, and find a KS test statistic of 0.25, but a p-value of 18.6\%. The AD test statistic is -0.69 with a p-value 61.2\%. Therefore, from our dataset, the LR and VR centaurs have indistinguishable eccentricity distribution.

\subsubsection{Resonant objects}
For resonant bodies, we get a KS test statistic of 0.12, and a p-value of 46.5\%. The AD test statistic here is -0.29 with a p-value of 85.3\%. 
{However note that, as seen in the Resonant objects panels of Fig. \ref{fig:eccMichael}, while the plutinos contain both LR and VR bodies, all objects beyond the 5:2 MMR are LR.} This implies that higher order resonances, that co-exist with the scattered disk, seem to contribute to the eccentricity-color correlation. As VR objects mostly have ecc$<$0.42, it is expected that the color-eccentricity correlation does not exist in resonances interior to the 5:2 because any object reaching e=0.42 in those populations would cross Neptune's orbit and be scattered on short timescales.
\subsubsection{Scattered disk objects}
Finally, we consider only scattered disk objects. In this case, the KS test statistic is 0.40, with a p-value of 17.5\%, and the AD test statistic is +1.18 with a p-value of 7.9\%. This implies that the LR and VR scattered disk samples have a 7.9\% or 17.5\% chance of being drawn from the same parent distributions, depending on the statistic test.

\subsubsection{Conclusions}
Put together, these statistics imply that the eccentricity distribution difference between the two color groups is caused by gray/LR objects strongly dominating in number in the scattering disk (and to a lesser extent, in the detached object population but they are less numerous in our sample), and in high-order resonances where e $>$ 0.42 is allowed. Moreover, this suggest that any Kuiper Belt formation model attempting to explain the data should simultaneously account for the lack of VR objects in the scattered disk, and their significant presence in the 3:2 MMR.

}

\section{N-body simulations}
\label{modelsec}
\subsection{Numerical setup}

In this section we interpret the observational results above using N-body simulations. The broad brush goal is to put forward a theoretical framework to interpret the dynamical origins of the color-eccentricity and color-inclination correlations, and use these to extract informations on the history of the solar system and the protoplanetary disk in which it formed.

All simulations in the work were done using the \texttt{REBOUND} N-body integrator \citep{r4,r2,r3}, along with its \texttt{REBOUNDx} add-ons package \citep{r1}. 
Simulations were performed using the \texttt{MERCURIUS} hybrid integrator that combines the fast and symplectic Wisdom-Holman integrator \texttt{WHFAST}, with the slower and non-symplectic but 15th-order precision \texttt{IAS15} integrator to efficiently resolve close encounters. This integration scheme is analogous to \texttt{MERCURY} \citep{1999MNRAS.304..793C}. We set $\Delta t = 0.5$ yr for \texttt{WHFAST}, while \texttt{IAS15} has adaptive timestepping that we limit to a minimum of $10^{-4}$ yr. In \texttt{WHFAST} we turn off the \texttt{whfast.safe$\_$mode} flag and turn on the symplectic correctors up to 11th-order for a significant increase in both accuracy and performance over the default settings. 

We set the critical radius inside of which the integration scheme switches the integrator from the fast \texttt{WHFAST} to the more accurate \texttt{IAS15} to a standard 3 Hill radii. Therefore, this is the radius at which the integrator assumes that a close encounter is taking place. We find that while smaller values lead to an unnacceptable loss of information, larger values have a significant numerical cost with diminishing return.
For Neptune at its current location, the critical radius is roughly 2.3 AU.  Moreover, we follow \cite{volk} in modifying \texttt{MERCURIUS} to exclude the planets from being flagged as undergoing close encounters. The integrator will hence always integrate the planets with \texttt{WHFAST}, and, during close encounters, switch to \texttt{IAS15} only for test particles. This ensures that close encounters with planets do not affect their orbital histories, allowing them to be bitwise reproducible and machine independent. {This modified version of \texttt{MERCURIUS} is available upon request.}

The system is initiated with the 4 giant planets on orbits summarized in table \ref{table1}. Jupiter and Saturn start on their current orbits, with their eccentricities and inclinations damped by a factor 2. Uranus and Neptune start significantly inward to their current orbits, and are then forced to migrate outwards on an e-folding timescale $\tau_m$ that we set to 60 Myr. {Migration is implemented using a new \texttt{REBOUNDx} custom force analogous to the pre-existing \texttt{modify}-\texttt{orbits}-\texttt{forces}. The new force, called \texttt{exponential}-\texttt{migration}, is now part of the standard \texttt{Reboundx} library, and is available for general use.} This force applies continuous velocity kicks to the particles :

\begin{equation}
\Delta v_{j}=\frac{1}{2} \frac{\Delta_{j}}{a_{j}} \frac{\Delta t}{\tau_m} e^{-t / \tau_m} v_{j}
\end{equation}

where $a_j$ is the semimajor axis at moment $j$, $\Delta_{j} = a_f - a_0$ is the migration distance, $v_j$ is the velocity at moment $j$, $\tau_m$ the migration timescale, and $\Delta t$ is the timestep. 

These continuous velocity kicks lead to the exponential change in the semimajor axis:

\begin{equation}
a(t)=a_{f}+\left(a_{0}-a_{f}\right) \times \exp \left(-\frac{t}{\tau_m}\right)
\end{equation}

The reasoning behind this simplified migration scheme is twofold. First, since our main goal is to gain an understanding into the origins of the eccentricity-color correlation, a simple model is easier to interpret and keep track of the relevant physical processes. Second, the additional processes modeled in the more complicated models are usually introduced to explain some of the finer details of the Kuiper Belt such as the abundance of bodies in individual resonances, and the properties of the ``kernel''. Since the number of objects in our observed dataset is limited, and detailed statistical analysis of individual populations is not possible due to untracetable observational biases being present in the colours sample, such additional model complexities are unlikely to allow for further significant insights at this point. 

\begin{figure*}
\begin{centering}
        \includegraphics[scale=0.25]{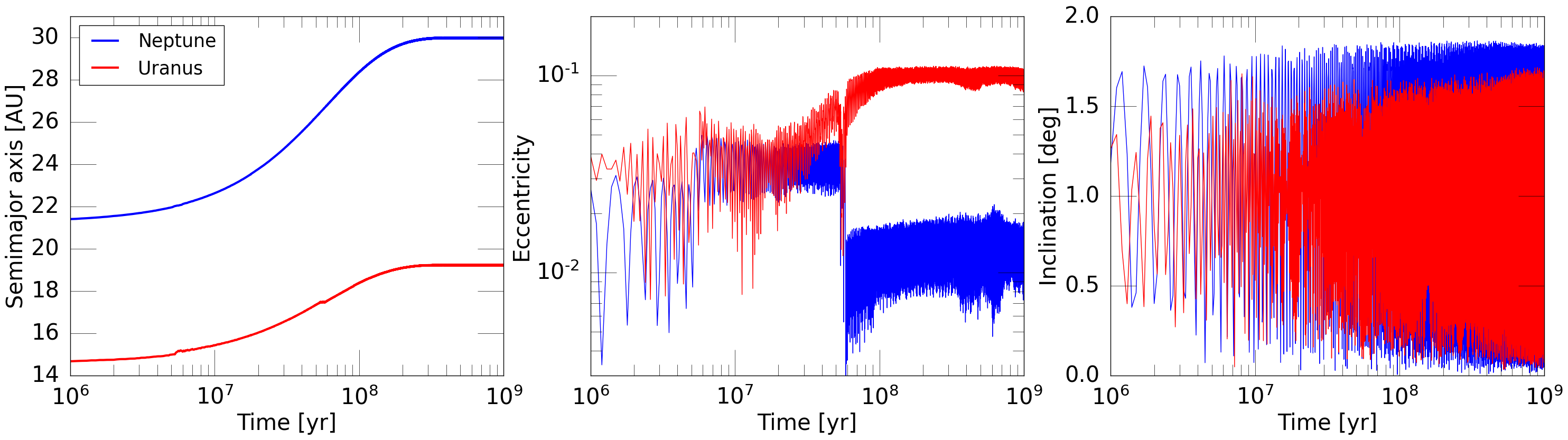}
   \caption{The orbital histories of Uranus and Neptune in our simulations. Planets are initiated as shown in table \ref{table1}, then migrated outwards on a 60 Myr e-folding timescale. {The eccentricity spikes around 50-60 Myr are caused by Uranus and Neptune crossing their mutual 2:1 MMR. } }
    \label{fig:planets}
    \end{centering}
\end{figure*}

{Our choice of $\tau_a$ (60 Myr) is motivated by the results of \cite{nesv2015} who suggested that the final inclinations distribution of TNOs is strongly correlated to Neptune's migration speed. However, as discussed in section 1, the recent results of \cite{volk} suggest that  the solar system's secular structure, mainly the giant planets' inclination secular modes f$_6$, f$_7$, and f$_8$, also contributes to this distribution. Since these modes are very sensitive to the planets initial eccentricities and inclinations, we follow \cite{volk} by fine tuning the initial conditions until we find an adequate set of dynamical parameters.} We hence run a large set of short simulations ($\sim$ 800 Myr) with just the 4 giant planets, where we fix all of the parameters for Jupiter and Saturn, while allowing small random perturbations to the initial semi-major axis (-0.1 $< \delta a < $ 0.1), eccentricity (0 $< \delta e_U < $ 0.048 and 0 $< \delta e_N < $ 0.0092), and inclination (0 $< \delta i_U \text{[rad]} < $ 0.013 and 0 $< \delta i_N \text{[rad]} < $ 0.03) for both Uranus and Neptune. We finally select the set of free parameters that lead to the most reasonable giant planets architecture (closest semimajor axis, eccentricity, and inclination to the observed values).  
Once these free parameters have been tuned, they are used in all of the different simulations, leading always to the exact same evolution of the planets due to the bitwise reproducibility of our integration scheme. The exact nominal values we use are reported in the Appendix. The orbital evolution of Uranus and Neptune in our simulations is shown in Fig. \ref{fig:planets}. At the end of the migration phase, Uranus' eccentricity and inclination are respectively within a factor of $\sim$ 2 and 1.25 of their current values. Neptune's eccentricity and inclination on the other hand are respectively within a factor of $\sim$ 1.6 and 1.7 of their current values. The final periods ratio of the two planets is almost identical to the measured value. 
We now calculate the secular frequencies amplitudes of Neptune in our simulations and compare them to reality  as suggested by \cite{volk}. We hence use linear secular theory as described in \cite{orangebook}. Neptune's secular inclination evolution is controlled by three modes: $f_6$, $f_7$, and $f_8$. The frequencies of these modes to first degree depend only on the semimajor axis and planetary masses. Since we use the exact masses and match the final semimajor axis very closely, we use the same frequencies as today's solar system with $f_6$ = -25.73355 arcsec/yr, $f_7$ = -2.90266 arcsec/yr, and $f_8$ = -0.67752 arcsec/yr. The modes amplitudes are then obtained by fitting the inclination's equation of motion:
\begin{equation}
I_N^2 = (\sum_{k=6}^{8} I_{N,k} \cos \left(f_{k} t+\gamma_{k}\right))^2 + (\sum_{k=6}^{8} I_{N,k} \sin \left(f_{k} t+\gamma_{k}\right))^2
\end{equation}  

For the dominant $f_8$ mode, we obtain: $I_{N,8}^{sim} / I_{N,8}^{obs} \sim 0.74$, with the other two amplitude the ratios slightly closer to unity.

Our test particles disk is initiated with a semi-major axis distribution $\propto a^{-2}$ ranging from 22 to 48 AU, a Rayleigh eccentricity distribution with a scale of 0.08, and a Rayleigh inclination distribution with a scale of 0.04 rad. 

We start the simulation with a total of 8600 particles. Since the capture and retention probabilities in some TNOs populations are very low, we use a particles cloning scheme to keep their number as high as possible.

At regular intervals (10$^3$ yr during Neptune's migration, then 10$^4$ yr afterwards) we stop the simulation and check if the number of particles has decreased (thus particles either escaped the domain edge at 500 AU or collided with a planet or the sun). In this case we then clone a random sample of the remaining particles, with the sample size automatically calculated to bring the total back to the initial particles number (8600). We ``thermalize'' the particles clones by adding a random perturbation with an order of $10^{-6}-10^{-5}$ rad to the mean anomalies. This scheme ensures that a sufficient number of particles will still be present at the end of the simulation, without the significant numerical cost of cloning the entire particles population at regular intervals. At the end of our nominal simulation, we found that 36\% of the initial unique particles population is still represented in the data, with the remaining 64\% being clones. Even though the entire initial particles parameters space is still represented in the final data, original particles starting inside 35 AU represents a small fraction of the total remaining particles. To check the validity of our cloning approach, we compare the resulting test particles dynamical distribution at the end of Neptune's migration to another simulation where we simply clone all of the remaining particles whenever any escape. We find that the final semimajor axis, eccentricity, and inclination normalized distributions are all very similar in the two cases, as KS tests indicate a less than 1\% probability that the 2 samples are drawn from different distributions.

Finally, to represent the particles' colors, we save each particle's initial semimajor axis, and then map it to a ``color'' parameter in the data analysis phase. Particles with an initial semimajor axis inside the color transition line are LR, and vice versa. Our model's \texttt{Rebound} Simulation Archives are available upon request, and allow for an exact reproduction of our entire simulations and results.

\begin{table*}[]
\begin{centering}

\begin{tabular}{l|c|c|c|c|}
\cline{2-5}
                                             & {Jupiter} & {Saturn} & {Uranus} & {Neptune} \\ \hline
\multicolumn{1}{|l|}{{a {[}AU{]}}}    & 5.17             & 9.49            & 14.69 + $\delta a_U$          & 21.41 + $\delta a_N$            \\ \hline
\multicolumn{1}{|l|}{{e}}             & 0.024            & 0.029           & 0.048   + $\delta e_U$         & 0.009   + $\delta e_N$          \\ \hline
\multicolumn{1}{|l|}{{inc {[}rad{]}}} & 0.0011           & 0.021           & 0.004  + $\delta i_U$          & 0.03 + $\delta i_N$             \\ \hline
\end{tabular}
   \caption{Giant planets initial conditions. Jupiter and Saturn are started on roughly their current orbits, while Uranus and Neptune are subsequently migrated outwards. }
      \label{table1}

   \end{centering}

\end{table*}

\subsection{Numerical results}

\subsubsection{Simulated orbits vs observations}
\label{orbitsresults}
Before we can discuss the origins of the color-eccentricity correlation, we need to make sure that our simulated Kuiper Belt reasonably reproduces observations in the space of orbital parameters. We hence use the OSSOS Survey Simulator 2.0 \citep{cold2,petit1,ban1,ban2,lawler} to bias our numerical data and then compare it directly to the observed OSSOS sample. We model the absolute magnitudes of the simulated objects by sampling the empirical two-slopes distribution found by \cite{fraser2014}. 





\begin{figure}
    \centering
    \subfigure{\includegraphics[scale=0.31]{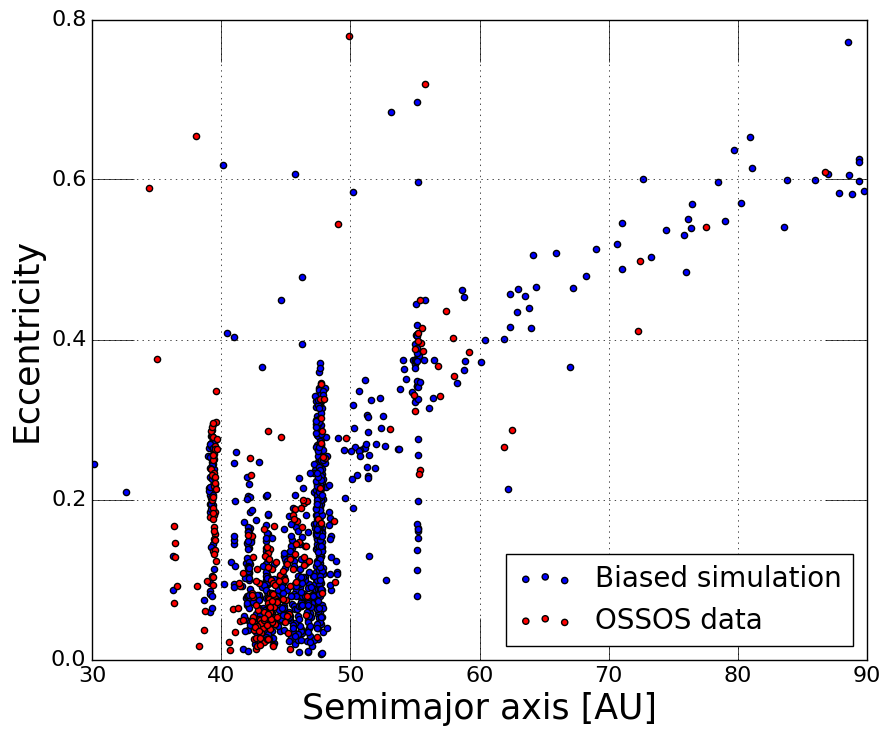}} 
    \subfigure{\includegraphics[scale=0.31]{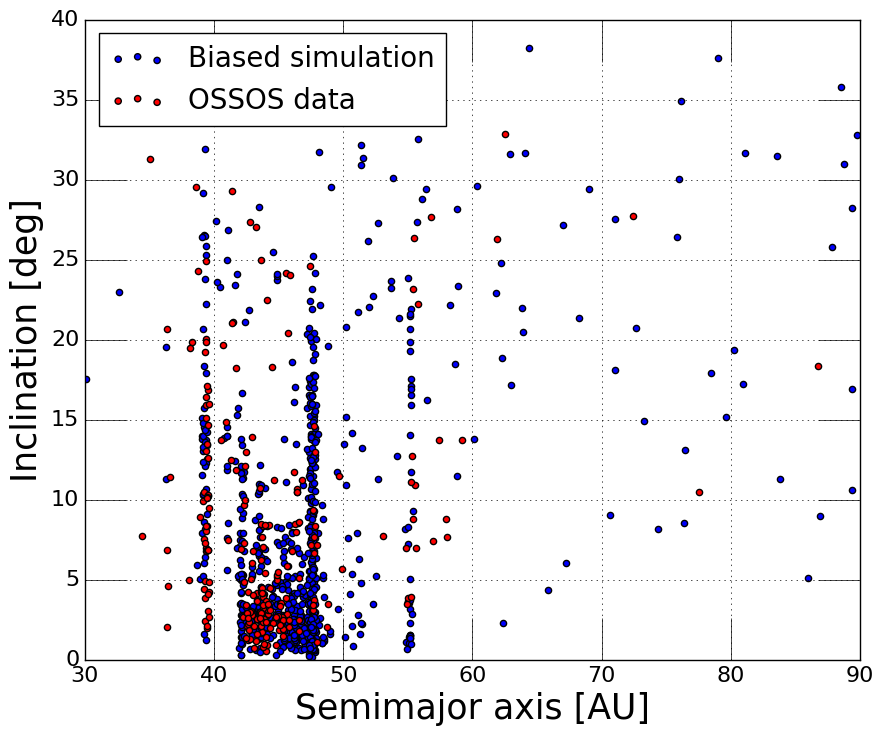}} 
    \subfigure{\includegraphics[scale=0.31]{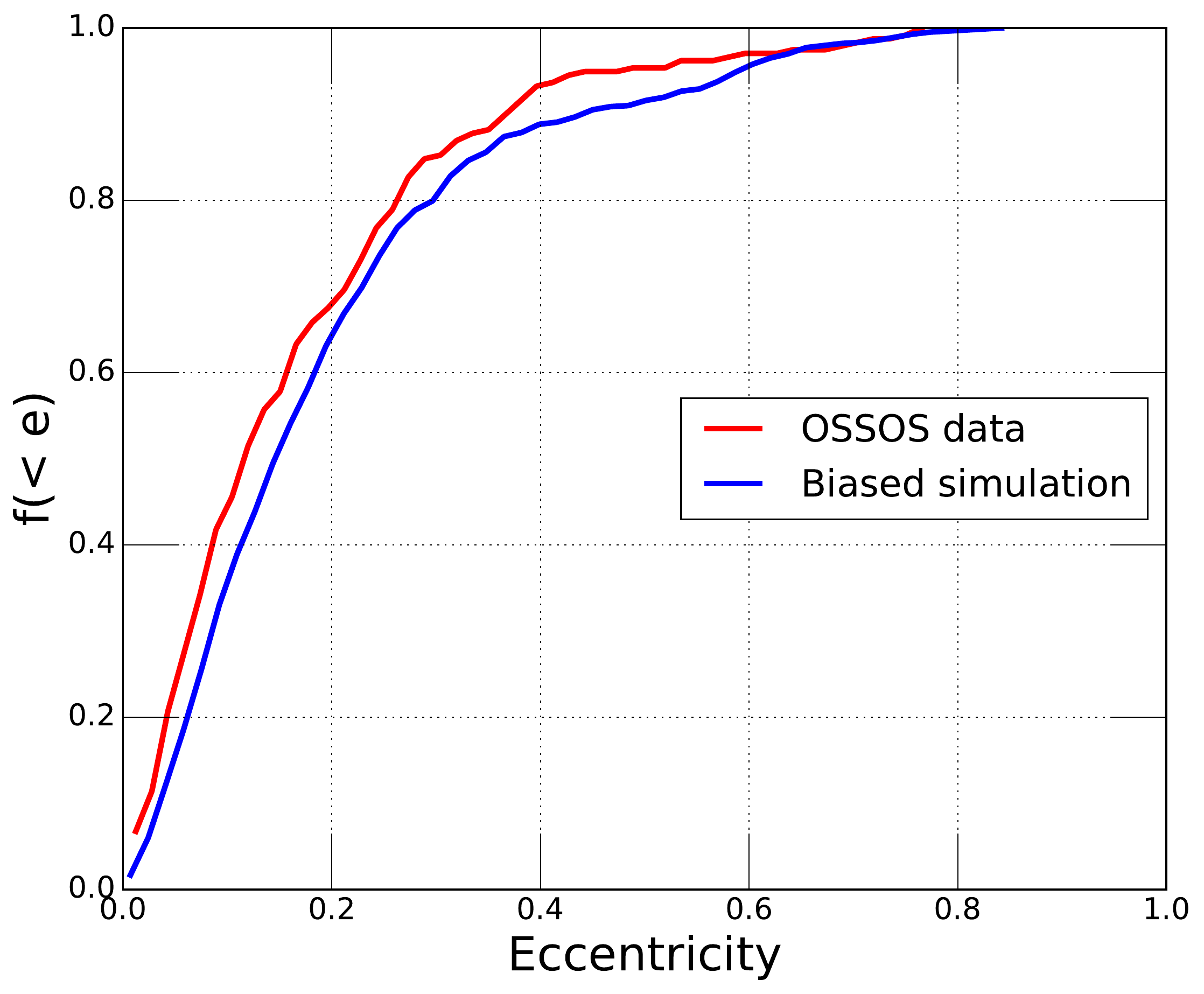}} 
    \subfigure{\includegraphics[scale=0.31]{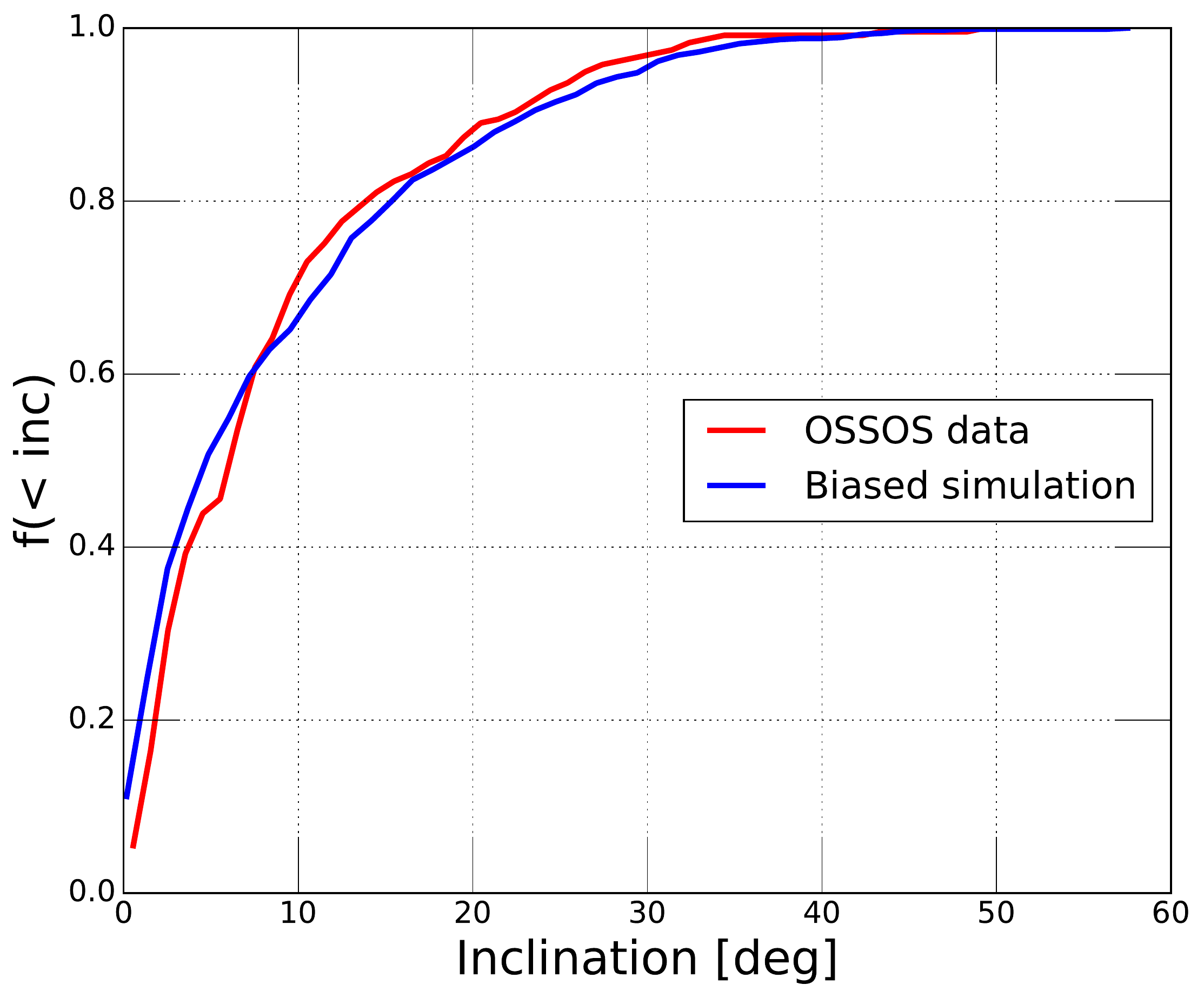}} 
  
    \caption{Top panels: the eccentricity and inclination distributions as a function of the semimajor axis for our observationally biased simulations results, compared to the OSSOS survey data. Bottom panels: Our biased eccentricity and inclination cumulative frequency distributions compared to OSSOS data.}
    \label{fig:aeinc}
\end{figure}

In Figure \ref{fig:aeinc} (top panels) we show the semimajor axis - eccentricity/inclination distribution of our nominal model post observational biasing using the survey simulator, compared to the OSSOS data. In Figure \ref{fig:aeinc} (bottom panels) we compare the eccentricity and inclination cumulative distributions of our model and the OSSOS data. Both figures show good agreement between our nominal model and observations. While our eccentricity distribution is slightly more low-eccentricity heavy than observations, possibly due to our choice for the particles initial semimajor axis distribution $\propto a^{-2}$, our inclination cumulative distribution matches observations well.  { Note that, during testing, we found the slow migration timescale to be important in reproducing the inclination distribution, as proposed by \cite{nesv2015}. }. Performing 2-sample A-D test, we find a 97.5\% chance that the two samples are drawn from the same parent distribution.
We obtain similar value for the eccentricities distributions, but only if we exclude $e<0.05$, as this is where the difference between the two plotted cumulative curves originates. Our model however is clearly imperfect, as for example it predicts significantly more objects in the 2:1 MMR than observed. {We note that this is probably a consequence of using smooth migration combined with a shallow disk profile. This might not be a  problem if for example disk profile is steeper and Neptune's migration is grainy.}

It is important to emphasize that we are not aiming to reproduce individual populations of the Kuiper Belt (such as the ratio of particles between specific resonances, or the abundance of trailing particles), but rather for a reasonable global sketch. This is because the dataset we are trying to explain is not large enough to allow for statistically studying individual populations. textbf{ Moreover, our model also does not aim to fit the cold classical population precisely, as for example we did not include a jump in Neptune's semimajor axis during its migration, which was shown to be important in preserving the ``kernel'' of the cold belt \citep{nesv2015a}.  }


\subsubsection{Color distributions: transition line location and color ratios}
The origins of the TNOs' surface colors are still far from being understood. In this work we investigate the possibility that the colors are primordial as is implied from observations \citep{fraser2012}, and thus set by the local physical-chemical structure of the protoplanetary disk when the objects formed. We assume a priori that the objects colors depend entirely on their formation location. There hence must be an abrupt chemical or physical ``color transition line'' in the disk that separated the formation zones of the LR and VR subpopulations. Tracing back the exact location of this transition line from the colors of TNOs is complicated by the fact that both the color transition line's location and the details of Neptune's migration will affect the mixing ratio of the two colors. Breaking this degeneracy necessitates knowing in advance either the transition line location, and using it to constrain the dynamics, or the other way around.

As our results in section \ref{orbitsresults} reproduced reasonably well the maximum eccentricities and inclinations of the observed dynamically excited objects, we will fix our dynamical model and vary the color transition location. We emphasize however that the solution we find is probably not unique, and assuming other early dynamical models might lead to different results.

Since our observational sample is limited, we will only fit the general aspects of the color distributions: mainly that VR objects are predominantly limited to eccentricities and inclinations below ecc=0.42 and i=21 deg.

\begin{figure}
    \centering
    {\includegraphics[scale=0.23]{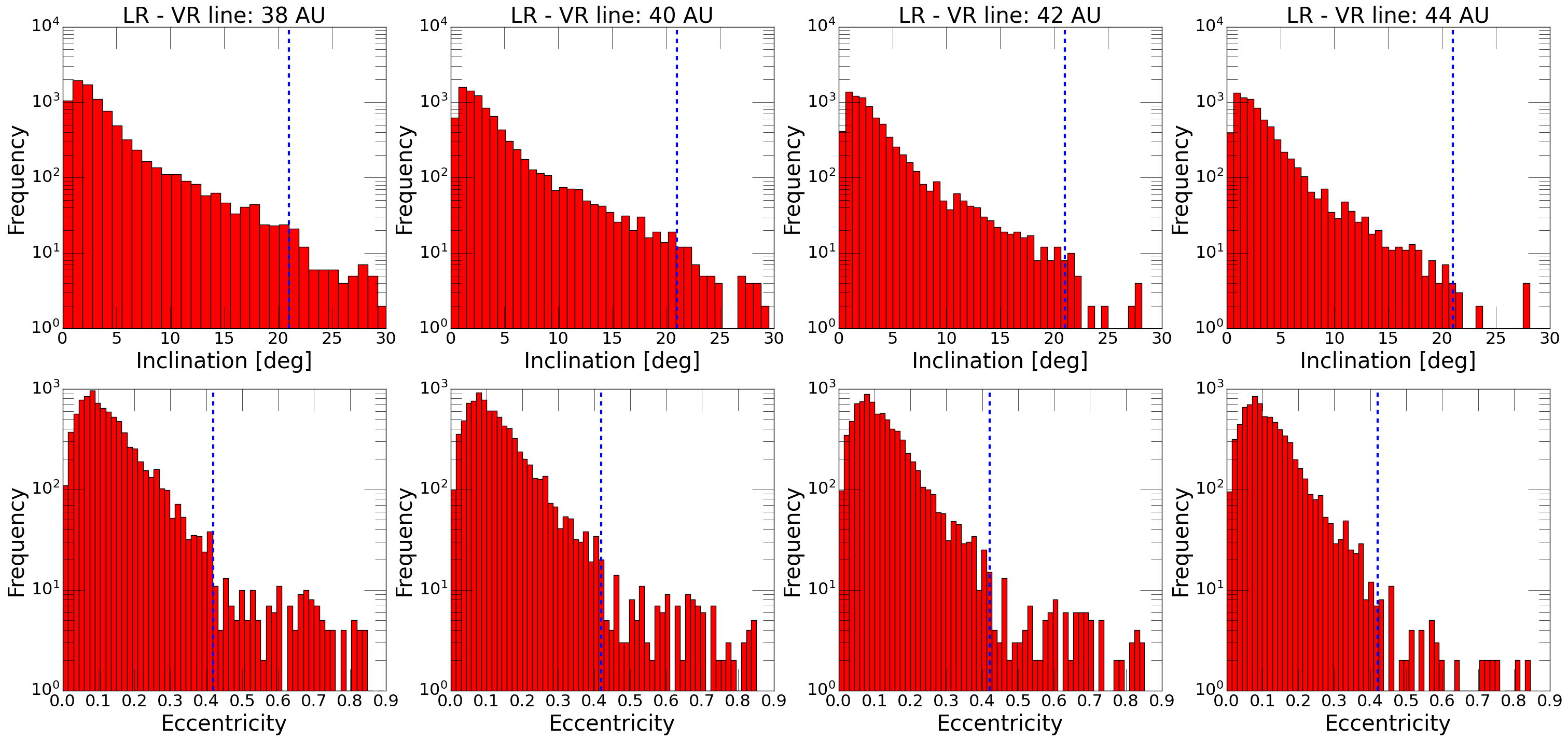}}

    \caption{Top panels: Final inclinations frequency histograms of VR particles in our nominal model, for different LR-to-red transition locations. Here we exclude VR particles with eccentricities higher than 0.42. 
            Bottom panels: Final eccentricities frequency histograms of VR particles in our nominal model, for different LR-to-red transition locations. Here we exclude VR particles with inclinations higher than 21 deg.}
    \label{fig:chemlinepos}
\end{figure}

We constrain the location of the color transition line using two different approaches. First, we incrementally move the transition line while checking for a VR objects cutoff in eccentricity-inclination space, to find the transition location that brings the cutoff as close as possible to e=0.42 and inc=21 deg. In Fig. \ref{fig:chemlinepos} we plot the eccentricity and inclination frequency histograms for the VR particles in our nominal model, for three different transition line locations (40, 42, and 44 AU). Note that a color transition line at or beyond 44 AU is observationally excluded due to the VRness of the classical cold belt objects. We nonetheless include it for comparison. When considering inclinations we exclude particles with e $>$ 0.42, and for eccentrities we exclude particles with inc $>$ 21 deg. This allows us to independently constrain the color transition line based on the two different parameters, without cross-contamination, in a similar way to how we have analyzed the orbit distributions of the observed colours sample.

Inspecting the eccentricity plots (bottom panels), we find a change in the distribution's slope (a ``plateau'') around e $\sim$ 0.4 for all three transition line locations. Moving the line outwards only decreases the number of VR particles with e $>$ 0.4 (so VR SDOs), without changing the plateau value. Comparing the ratio of VR objects with eccentricities higher and lower than 0.42 with observations is hence needed to constrain the location of the color transition line using only eccentricity data. Unconstrained observational biases in our dataset preclude us from doing that comparison. Hence transition lines beyond 40 AU are all consistent with  the eccentricity observations, even though if we take the data at face value then a transition line at or beyond 42 AU is preferred. 

Inspecting the inclinations distributions (top panels), we find that for a transition line at 42 AU there is a sharp cutoff in the distribution around 23 deg. This cutoff moves toward higher inclination values then disappears for transition lines below 42 AU. Moreover, while it also moves toward lower inclination values for transition lines beyond 42 AU, we exclude such values as mentioned above. Therefore, we conclude that in our nominal model, the strong limits on VR objects in e-inc space is best reproduced by a color transition line at $\sim$ 42 AU.

A complimentary line of evidence that we can possibly use to constrain the location of the primordial color transition line is the LR-to-VR color ratio of objects, inside and outside of the {\color{black}Q4} quadrant. This method can be more objective and precise than our first method. It is however strongly plagued by the uncertainties in the observational biases of the dataset, in addition to the limited number of objects outside of the Q4 quadrant. 

While we will hence use the LR to VR ratios as rough guidelines, and emphasize that observationally  these ratios are biased \citep{marsset}. This analysis hence serves as a proof of principle for future studies with more data and fully characterized discovery biases, rather than a definitive result on its own.  { For the observed and simulated ratios to be directly comparable, we first pass the simulations through the OSSOS Survey Simulator, but use an albedo of 12\% for the VR objects, and 6\% for the LR objects \citep{lacerda}. We use a diameters distribution based on the Jupiter Trojans, following \cite{nesv2020}. This is redone for each value of the color transition line. We additionally follow M2019 in cloning all particles 10 times and randomizing their positional angles before feeding them to the Survey Simulator, to improve the statistics. Note while this introduces observational biases to the simulations, it does not alleviate the biases in the M2019 data discussed above. Moreover, we also caution the reader that the OSSOS survey simulator should ideally be used for surveys with a recorded pointing history and a characterised detection efficiency. Not all objects in the M2019 dataset have this information available. As a consequence, using the survey simulator to compare our simulated TNO population to the observed M2019 population has its limits. For instance, some surveys like OSSOS were designed to sample specific dynamical populations of TNOs (mostly the Plutinos in the case of OSSOS, see Fig. 1 of \cite{ban2}). So, by using the survey simulator without knowing the survey pointing history, one may artificially increase or decrease the number of simulated objects from a specific dynamical population. If this population has a distinct color ratio compared to other dynamical groups, this will introduce a color bias in the simulated population.}

\begin{figure}
    \centering

    \subfigure{\includegraphics[scale=0.31]{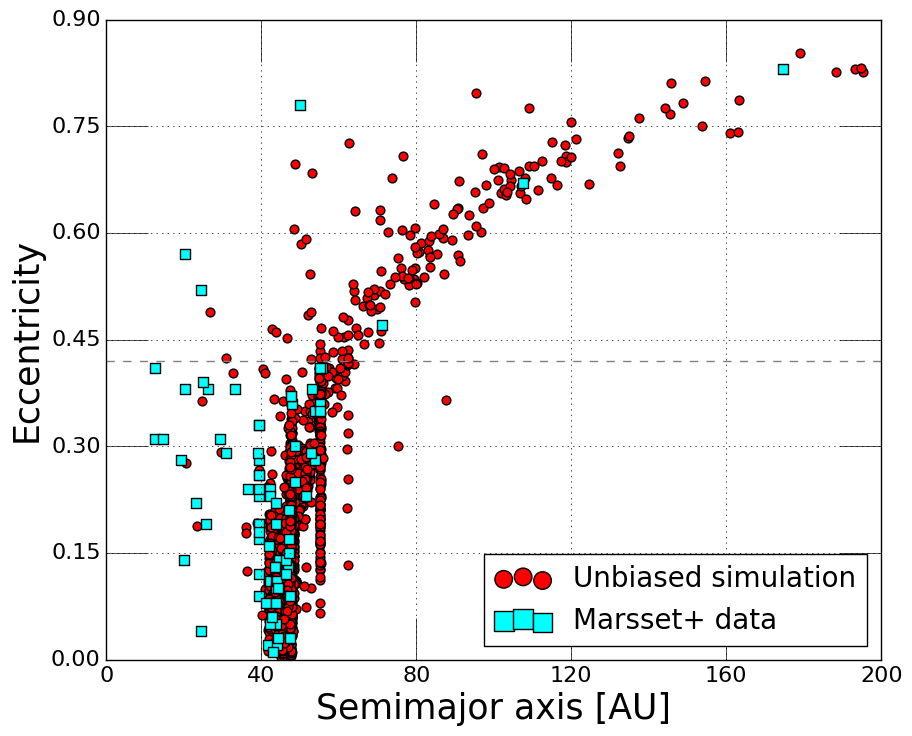}} 
          \subfigure{\includegraphics[scale=0.31]{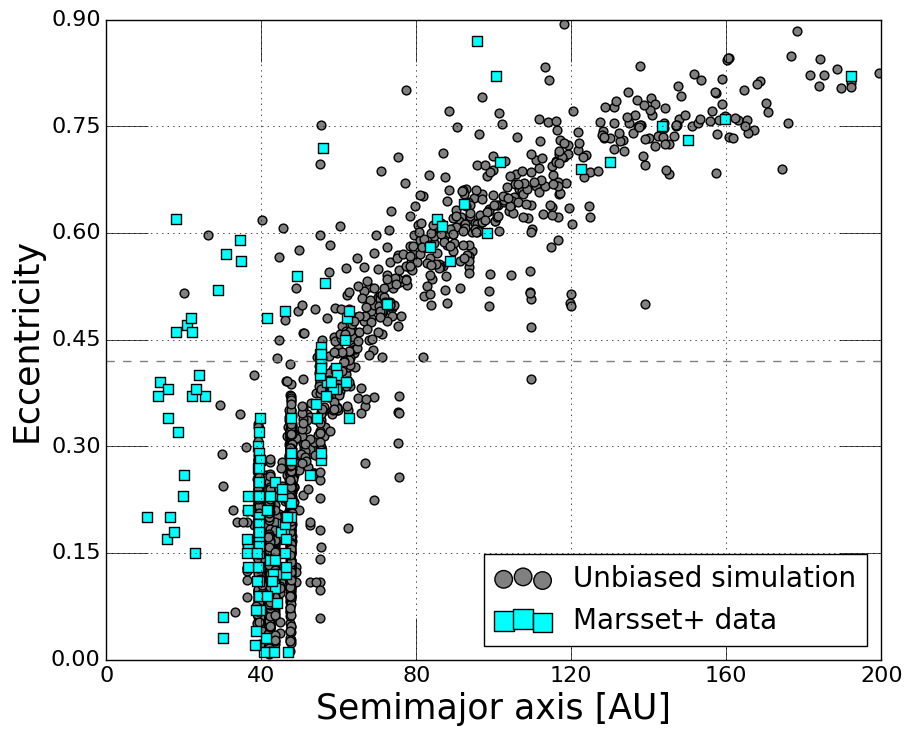}} 
    \subfigure{\includegraphics[scale=0.31]{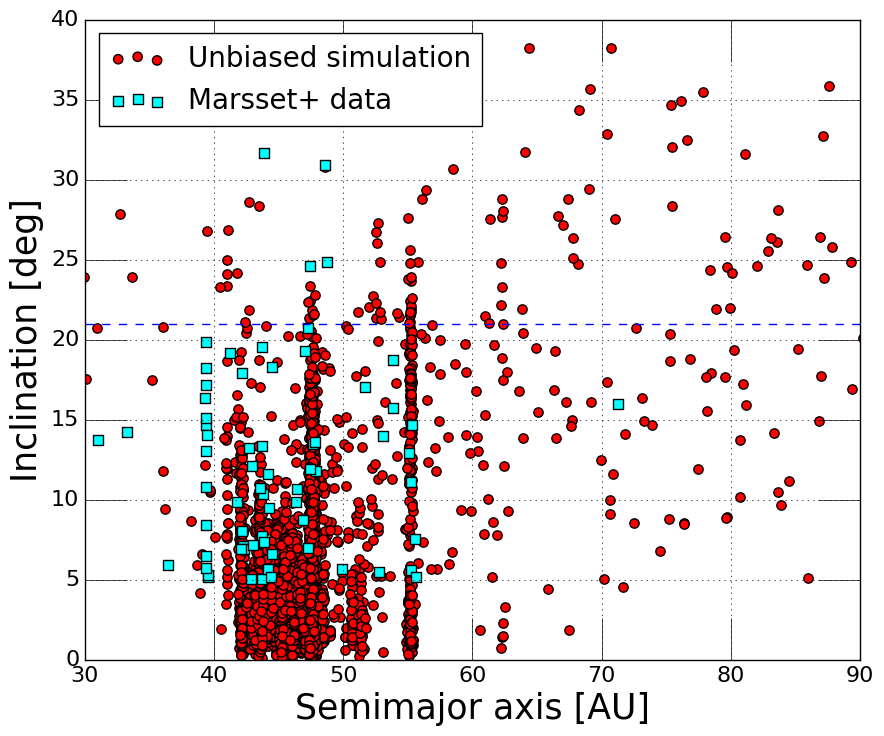}} 
        \subfigure{\includegraphics[scale=0.31]{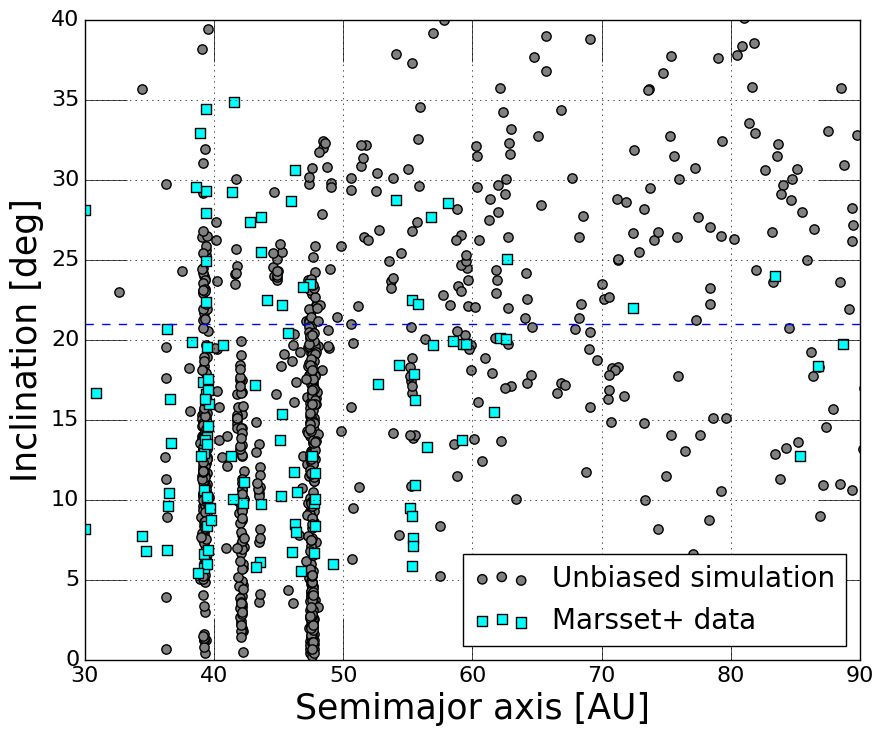}} 
    \subfigure{\includegraphics[scale=0.31]{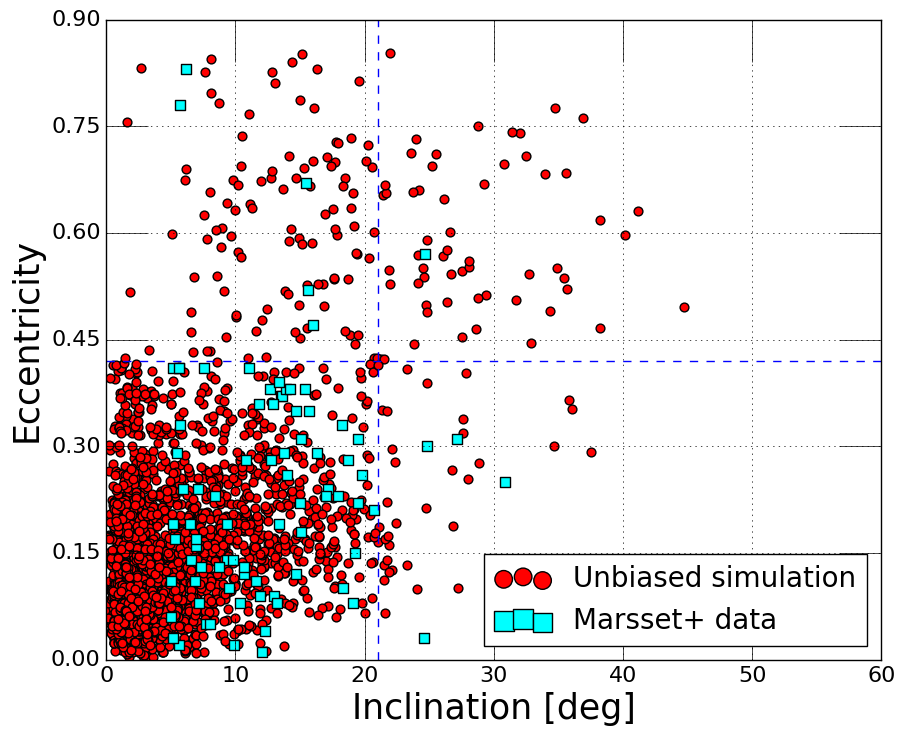}} 
    \subfigure{\includegraphics[scale=0.31]{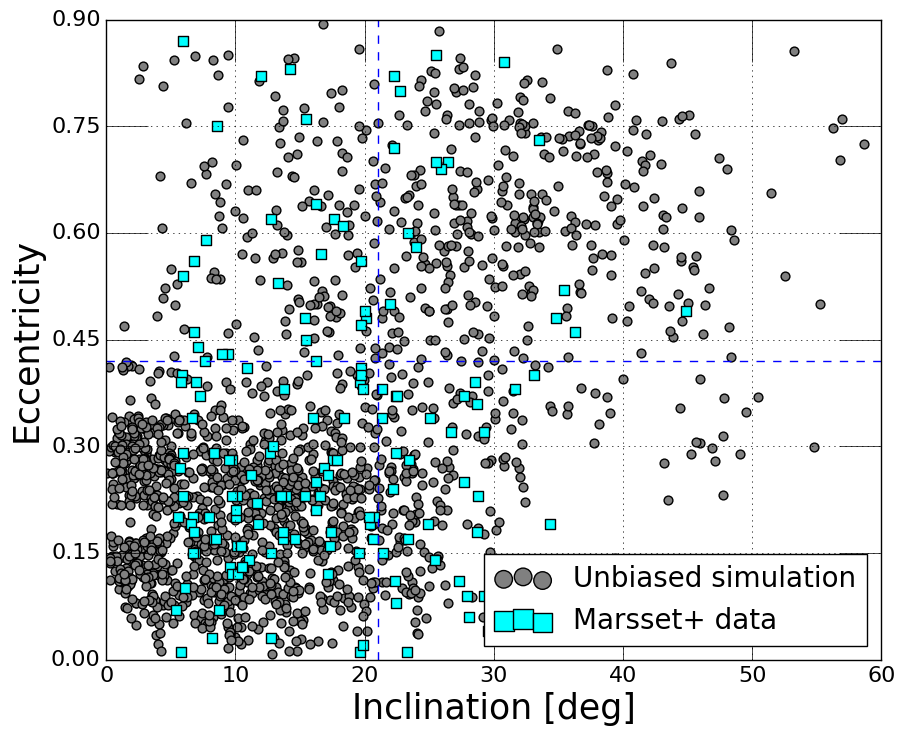}} 
        \caption{Left hand panels: Very red objects semimajor axis, eccentricity, and inclination diagrams from our full nominal  model results, compared to the M2019 dataset.
            Right hand panels: Same as above, but for red (gray) objects.  We emphasize that the M2019 dataset, by construction, does not include objects with inc $<5$ deg.}
    \label{fig:aeincUnb1}
\end{figure}

\begin{figure}
    \centering

      \subfigure{\includegraphics[scale=0.31]{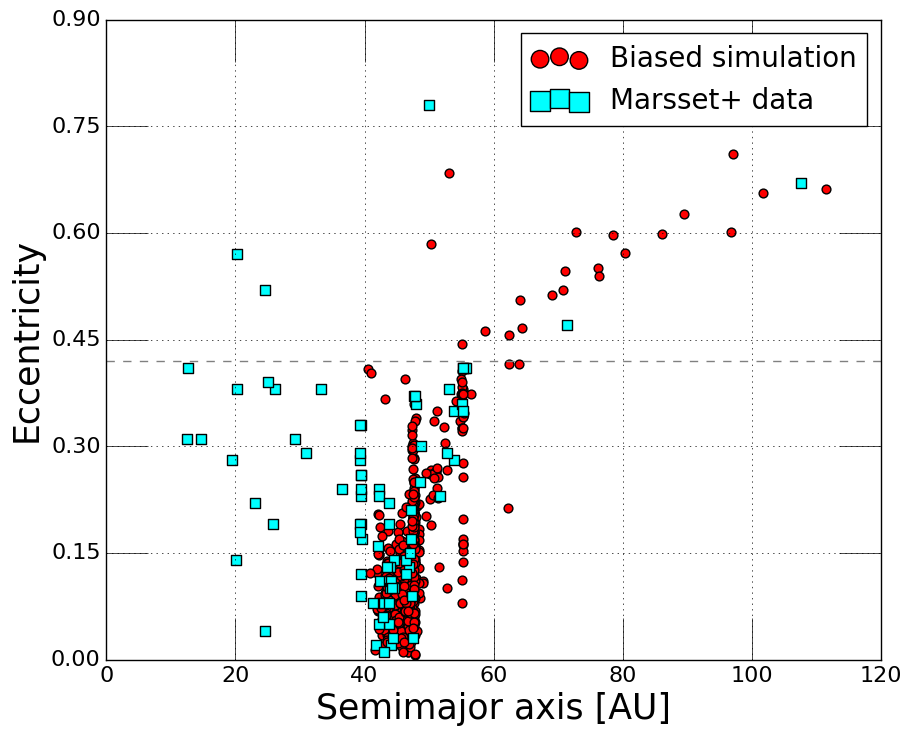}} 
            \subfigure{\includegraphics[scale=0.31]{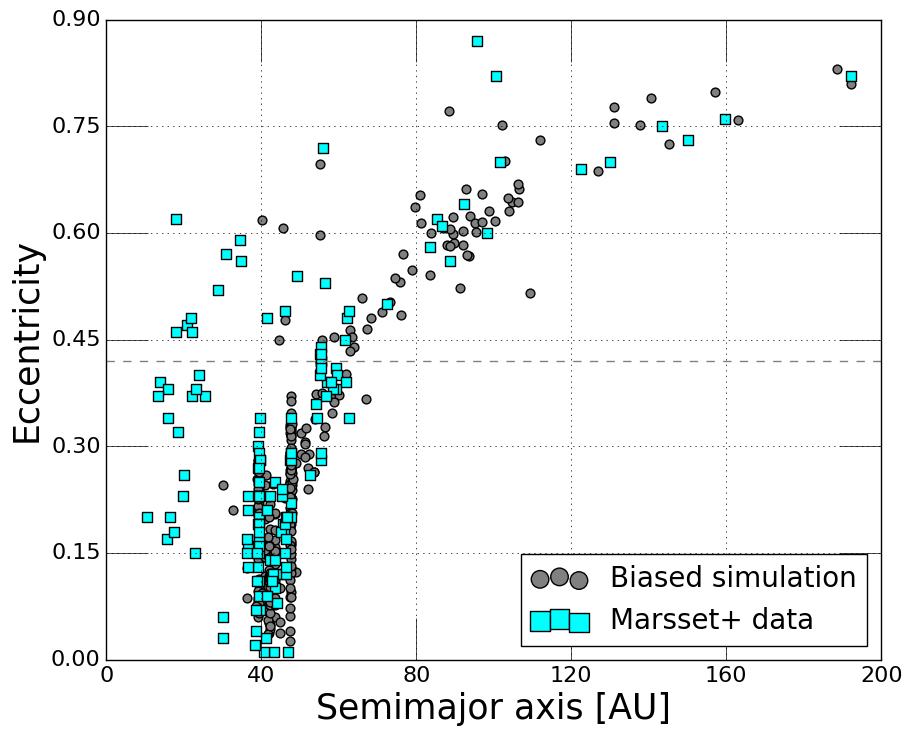}} 
    \subfigure{\includegraphics[scale=0.31]{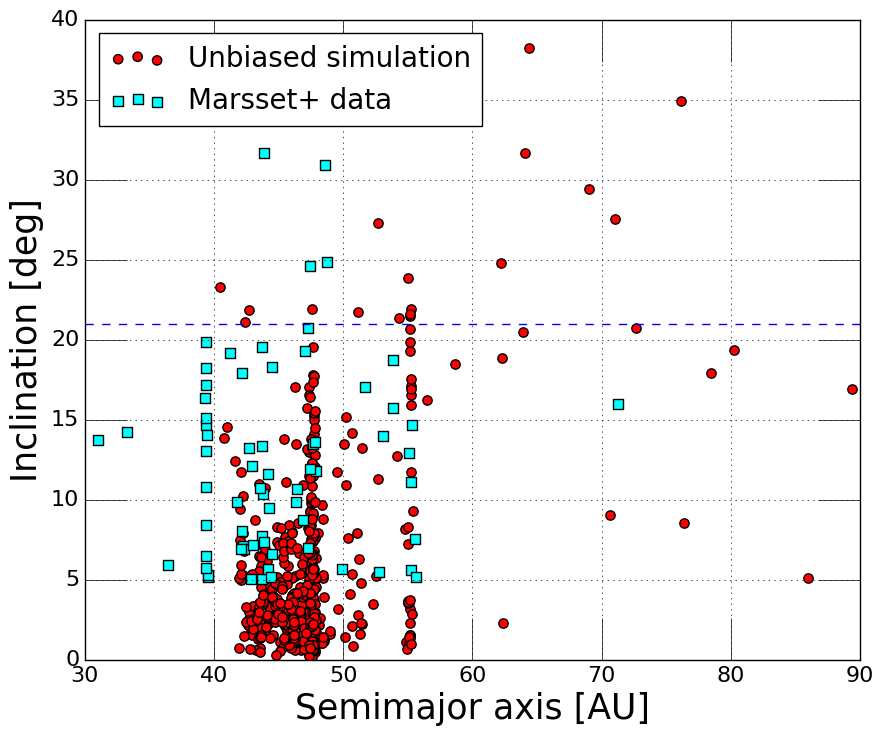}} 
        \subfigure{\includegraphics[scale=0.31]{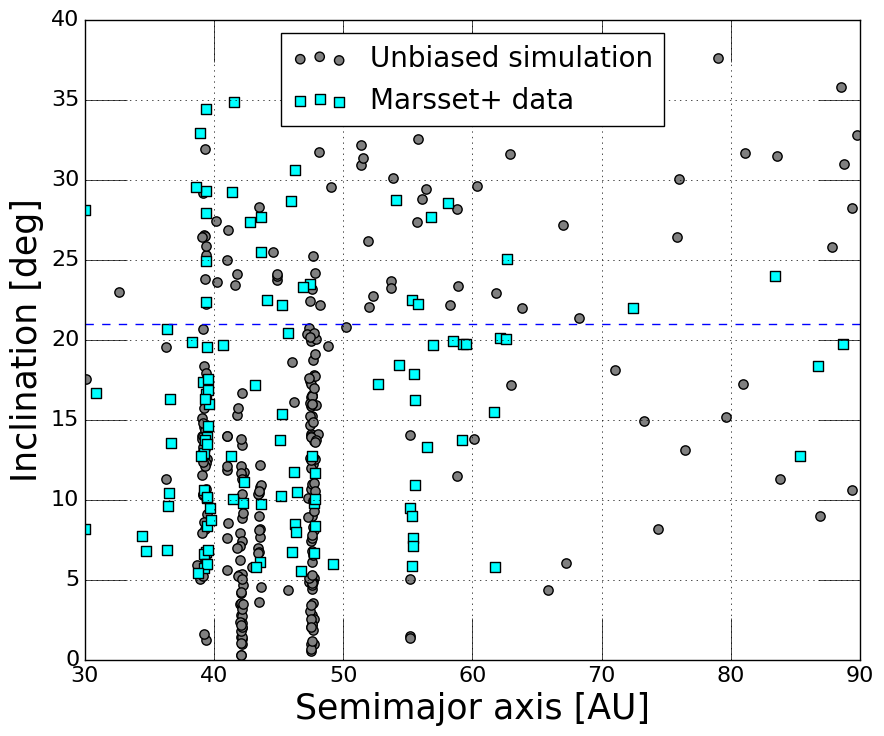}} 
    \subfigure{\includegraphics[scale=0.31]{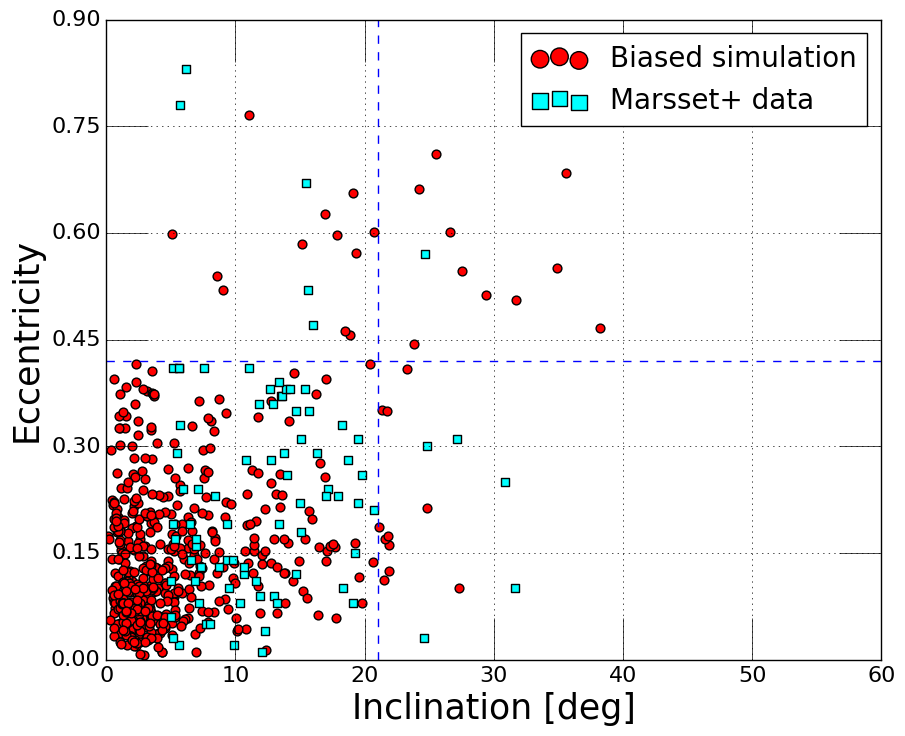}} 
    \subfigure{\includegraphics[scale=0.31]{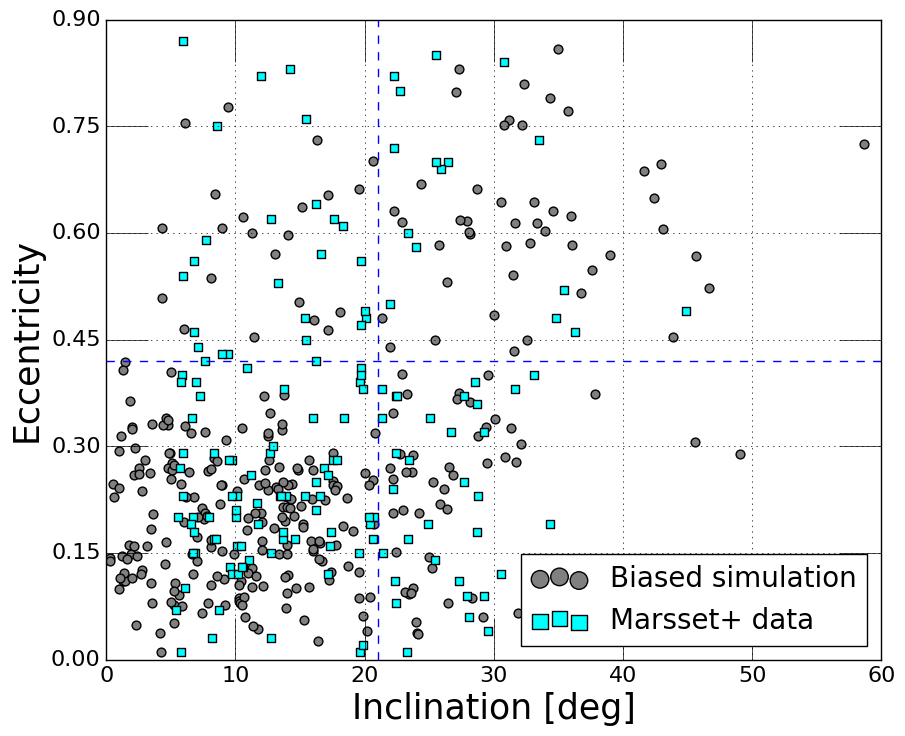}} 
                    \caption{Left hand panels: Very red objects semimajor axis, eccentricity, and inclination diagrams from our observationally biased model results, compared to the M2019 dataset.
            Right hand panels: Same as above, but for red (gray) objects. }
    \label{fig:aeincUnb2}
\end{figure}

{By gradually moving the transition line, we again find a best fit color transition line around 40-42 AU, with lines inside of 40 AU leading to a LR to VR ratio that is possibly too small outside of the Q4 quadrant. Our results, compared to observations, are plotted in Figs. \ref{fig:aeincUnb1} and \ref{fig:aeincUnb2}, for the unbiased and biased models respectively. For our nominal model (smooth migration for Neptune, and a planetesimals disk with $\Sigma\propto 1/r^2$) we choose a transition at 42 AU, and find a LR/VR ratios of 0.27 and 2.0 inside and outside of Q4, respectively. This compares decently with the observations if we take them at face value, where LR-to-VR ratios are 1.08$\pm$0.24 and 6.94$\pm$2.97 inside and outside of Q4, respectively. The intervals are 1-sigma Poisson error bars. For a transition line at 40 AU, these factors increase to respectively 0.18 and 0.63. For a transition at 35 AU, our model produces a VR-to-LR ratio 1 to 2 orders of magnitude lower than observations. These two figures (\ref{fig:aeincUnb1} and \ref{fig:aeincUnb2}) however also reveal a shortcoming of our model: the lack of VR objects in the 3:2 MMR. While the observational dataset contains $\sim 10$ such objects, our model produces virtually none.} This issue is also found in the grainy migration and cold disk models discussed later. This can only be rectified if the color transition line is at or inside of 38 AU, leading to a much larger number of red objects in the scattering disk. This tension between the observed VR objects in the scattering disk and those in the 3:2 MMR is not trivial to interpret theoretically, as a perfect model should be able to both populate the 3:2 MMR with almost an equal number of VR and LR objects, while virtually not scattering any VR objects into e$\geq$0.42, or capturing any in the higher order MMRs.

We hence conclude that a color transition line at either 38 or 42 AU is partially consistent with current observations, and that more modeling and data are needed to resolve this paradox. We hence choose a more conservative color transition line at 42 AU for the rest of this paper, as any VR objects in this case will still be VR for more close-in transition lines.  

Note that when calculating the ratio inside Q4 for our simulations, we exclude objects with inclinations less than 5 deg to be consistent with the observational sample. If we include these objects, our {\color{black}LR-to-VR} ratio inside Q4 decreases to 0.41. This significant change is due to cold classicals being almost entirely VR in our simulations, in agreement with observations.



\begin{figure}
    \centering
    {\includegraphics[scale=0.25]{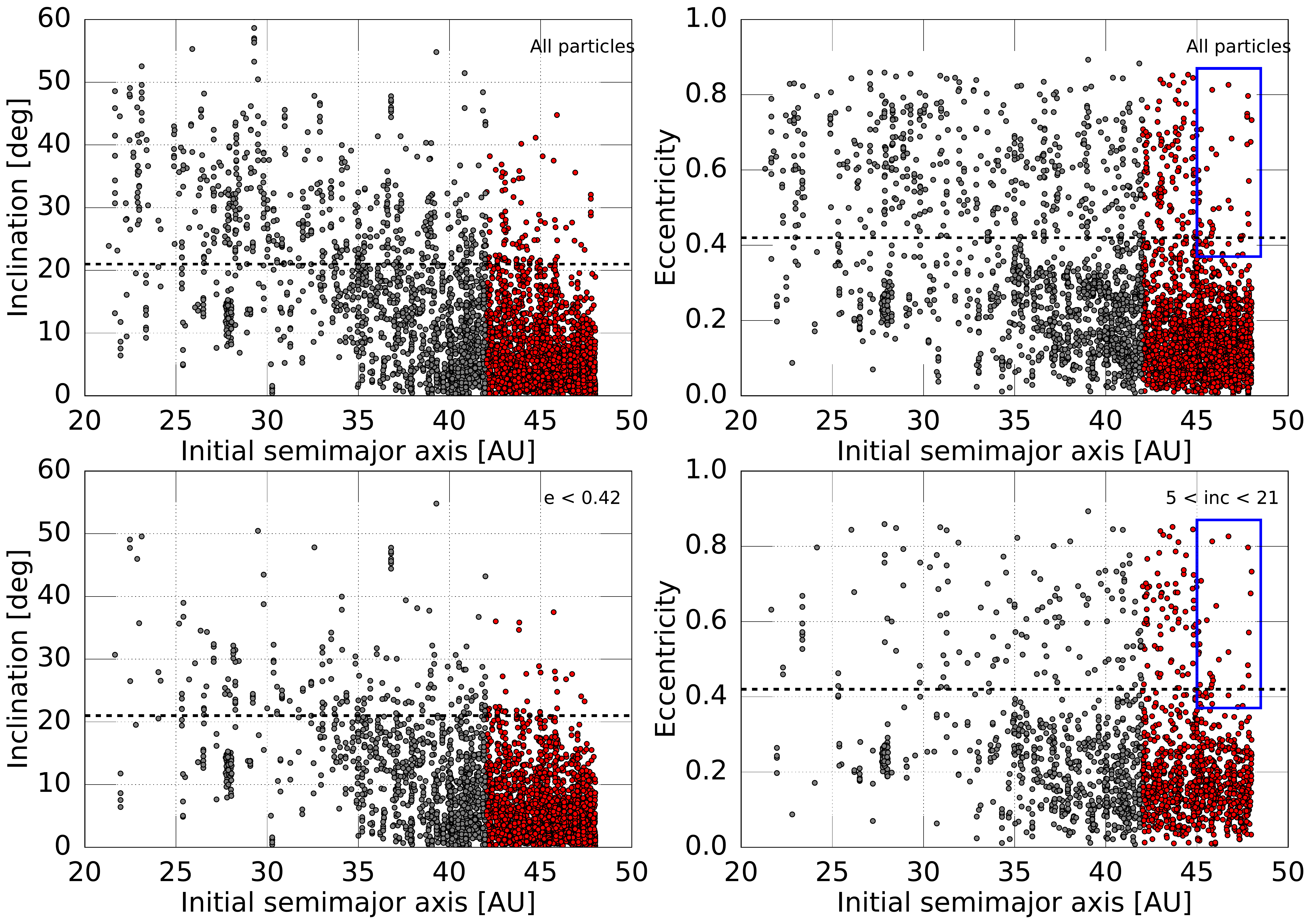}} 

    \caption{Top panels: the final eccentricities and inclinations for all of our simulated particles as a function of their initial semimajor axis. The blue box on the right panel delimits a zone with a paucity of high eccentricity VR objects, probably due to the g7, g8, f8, and f8 secular resonances. 
    Bottom panels: same as above, but limiting the inclinations plot to eccentricities less than 0.42, and our eccentricities plot to inclinations between 5 and 21 deg.   }
    \label{fig:initiala}
\end{figure}

For a deeper understanding of these ratios we plot the final eccentricity and inclination of our particles as a function of their initial semimajor axis in Fig. \ref{fig:initiala}. The maximum inclinations reached for objects shows a strong negative correlation with their initial semimajor axis, as one would expect since particles initially closer to Neptune will get more dynamically excited. This negative correlation naturally leads to a smaller number of high inclination VR objects than LR objects. {Note that here we also notice that the model allows for the formation of the cold belt in situ, as particles starting beyond 42 AU are mostly undisturbed and maintain their initially low inclinations. However, as we did not fine-tune the model to reproduce the details of the cold classicals, we do not dive further into this population's properties in our simulations. }

The maximum eccentricities of objects on the other hand shows a much weaker negative correlation with the initial semimajor axis, being roughly constant across all initial semimajor axis.

A simple maximal eccentricity - initial semimajor axis correlation hence does not explain the paucity of high eccentricity VR particles. Inspecting the eccentricities of particles beyond 42 AU however reveals a strong ``cliff'' around 45 AU, inside of which VR objects are as eccentric as LR objects, but beyond it the high eccentricity part of the diagram is mostly empty. This absence of high eccentricity particles is located between 45 and 48 AU. 
This can be possibly due to a lack of sweeping MMRs in the area between 45 and 48 AU. In fact, only the 2:1 MMR passes through this area during Neptune's migration, as the 5:3 ends up at $\sim$ 42.3 AU and the 7:4 at $\sim$ 43.7 AU. The next significant MMR, the 5:2, starts at $\sim$ 47 AU before ending at 55.5 AU. As the 2nd and 3rd order MMRs have capture probabilities that scale  with eccentricity as $e^{-1}$ and $e^{-1/2}$ instead of the $e^{-3/2}$ found for 1st order MMRs \citep{hahn2005}, these might be more relevant to the formation of high eccentricity red objects than the 2:1 MMR.

Another possible interpretation for the sudden decrease in the maximal eccentricity of red particles around 45 AU can be found in the results of \cite{holman} who simulated the eccentricity distribution of particles as a function of their semimajor axis, and found a cliff outside of 42 AU. This can be attributed to the  f$_7$, f$_8$, g$_7$, and g$_8$ secular modes found inside 42 AU, and sharply disappear beyond it \citep{Knezevic}. These modes can significantly contribute to the particle's dynamical excitation, and their absence beyond 42 AU will lead to dynamically cold populations, the most prominent of which is the classic cold belt. 
The 42 AU secular structure transition location reported by \cite{holman,Knezevic} however is different than the 45 AU transition found in our simulations, but this can be attributed to the many differences in the models. This include for example Uranus and Neptune's migration, that was not included in the \cite{holman} models, sweeping the outer edge of the g8 resonance through the area before stopping at 48 AU. Moreover, as the solar system's secular architecture is very sensitive to the initial conditions and exact evolution of the 4 giant planets \citep{volk}, even mild differences between models can lead to significantly different secular architectures. What gives some credence to this interpretation is that, when considering the low inclination red objects in the bottom panels of Fig. \ref{fig:initiala}, we can notice that at 48 AU, the particles maximum eccentricity distribution is once again suddenly similar to that of the LR particles, with maximal eccentricities rising up to 0.8.  This is consistent with  \cite{holman} and \cite{Knezevic} who both reported a reemergence of the g8 mode at 48 AU. 
In this picture, the paucity of VR objects at high eccentricities is due to the narrowness of the initial semimajor axis interval that can lead to such particles, as its constrained by the LR-VR color transition line on one side, and the secular resonances cliff on the other.

\subsubsection{Color distributions: Individual populations}
\label{indptheo}
The previous section gave us a bird's eye view for the origins of the color eccentricity and inclination correlations by linking the maximal eccentricity and inclination distributions to the initial semimajor axis of the objects. However, this does not tell us \textit{how}, physically, these high eccentricity or inclination VR objects reach their orbits. In this section we hence inspect the dynamical history of individual particles in our simulations. Our aim is mainly to understand the origins of the VR objects in the predominantly LR high eccentricity or inclination populations.

\begin{figure*}
    \centering
    \includegraphics[scale=0.4]{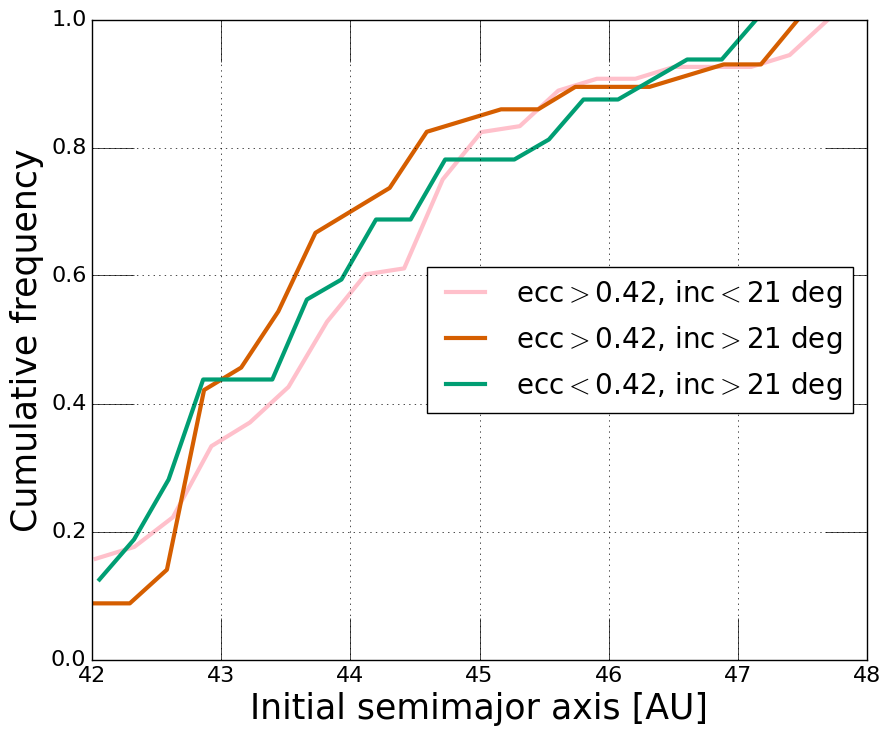}

    \caption{Normalized cumulative frequency distribution of the initial semimajor axis for the dynamically excited (Q1, Q2, and Q3) VR particles at the end of our nominal simulation. As we set our color transition line to 42 AU, this is the minimum value in the diagram.
    We see that all three subpopulations have similar initial location distributions even though, as argued in section \ref{indptheo}, they evolve into their characteristic orbits via different mechanisms. }
    \label{fig:freqchem}
\end{figure*}

We start by examining Fig. \ref{fig:freqchem} showing the initial semimajor axis of all of the outside of the {\color{black}(Q4)} VR particles (with either high eccentricity or inclination). We notice that the initial location of VR particles have no strong correlation with the final position of the particle in the e-inc diagram.

\begin{figure*}
    \centering
    \subfigure{\includegraphics[scale=0.15]{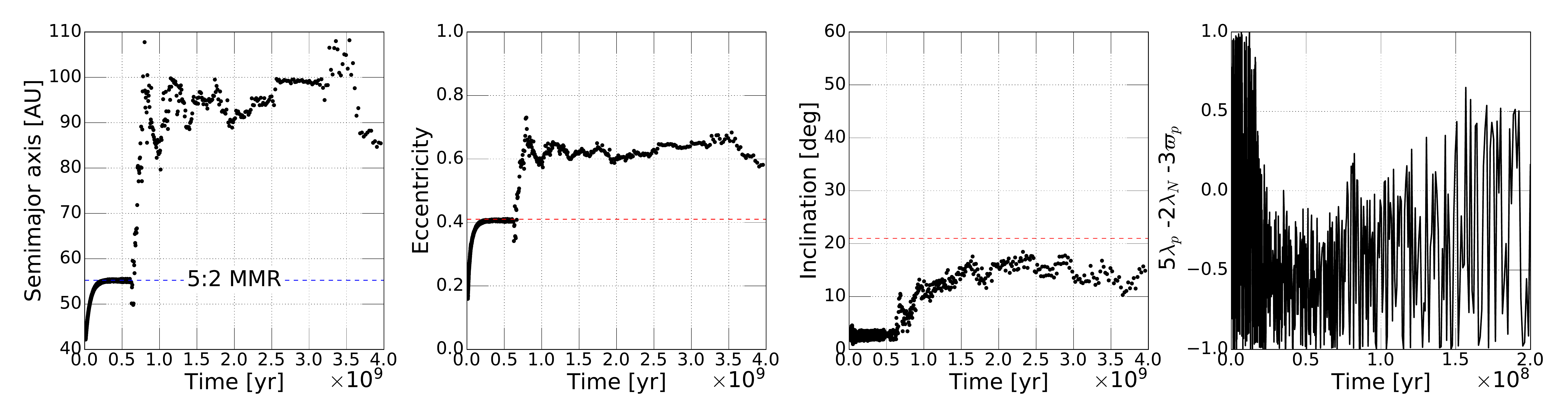}} 
    \subfigure{\includegraphics[scale=0.15]{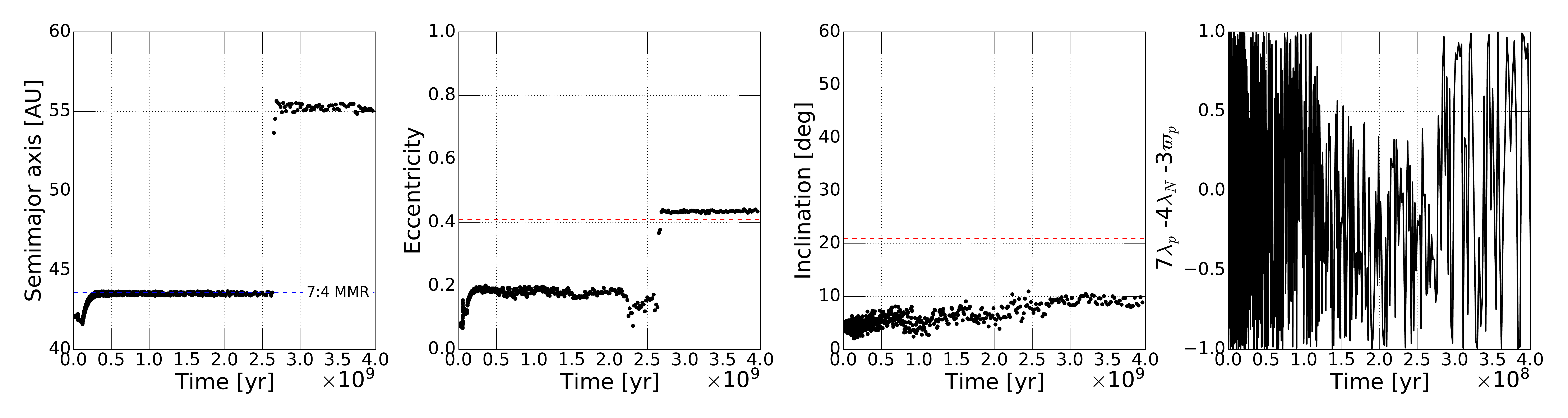}} 
    \subfigure{\includegraphics[scale=0.15]{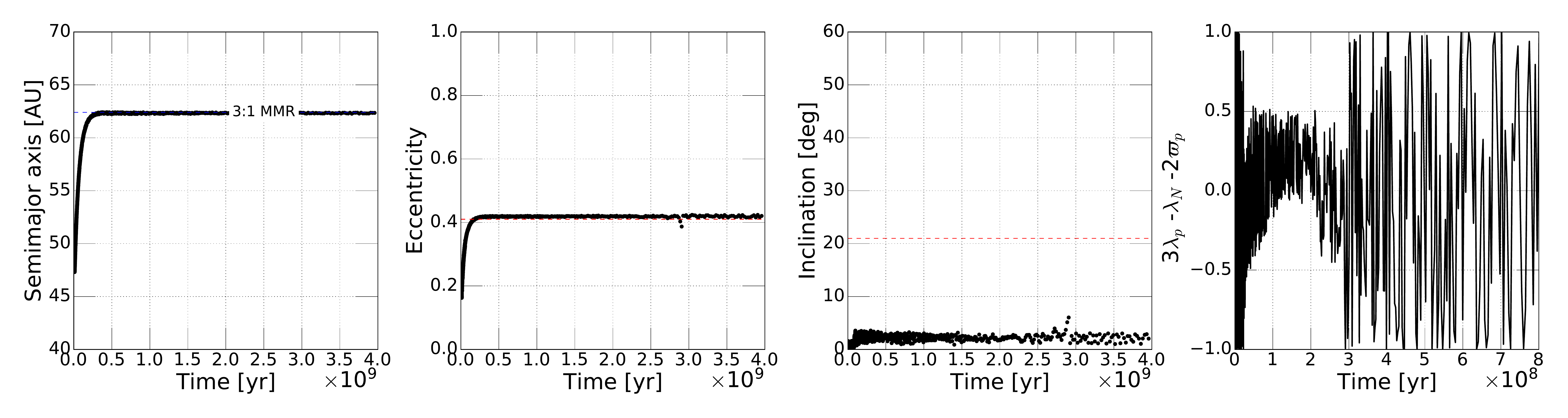}} 
    \subfigure{\includegraphics[scale=0.15]{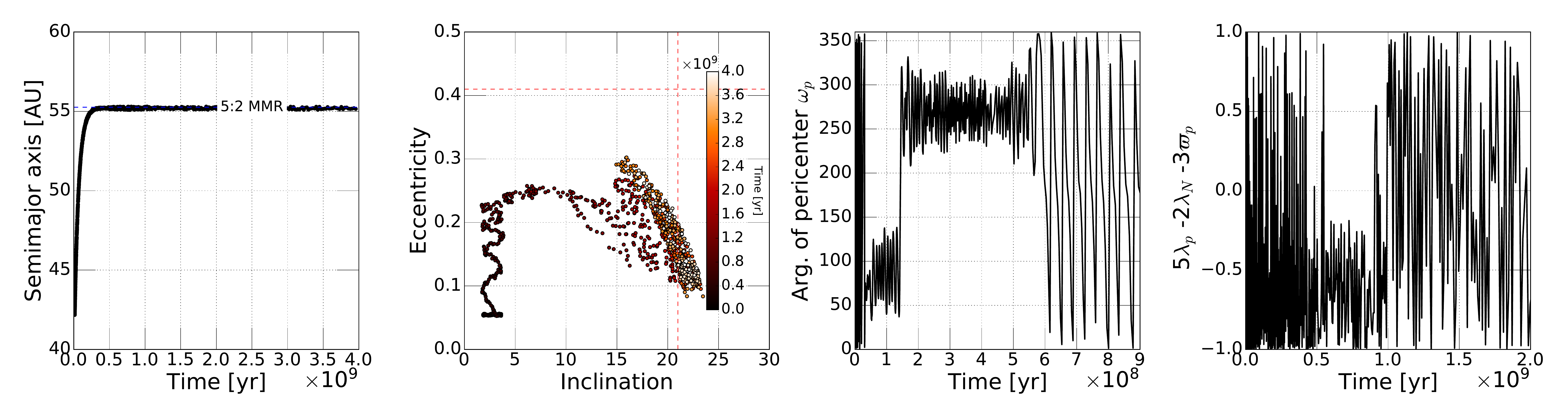}} 
    \subfigure{\includegraphics[scale=0.15]{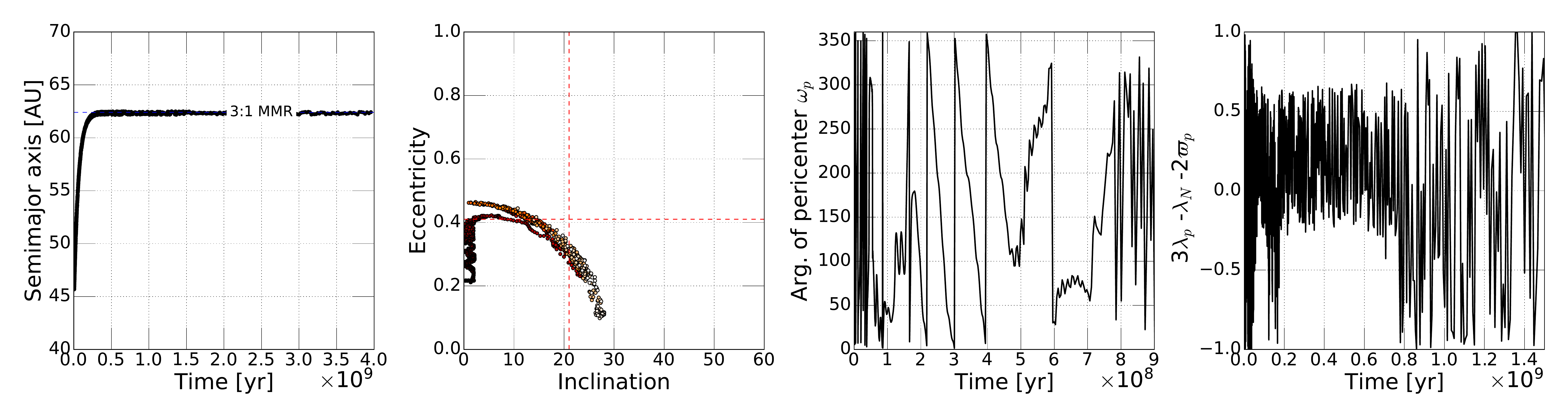}} 
    \subfigure{\includegraphics[scale=0.15]{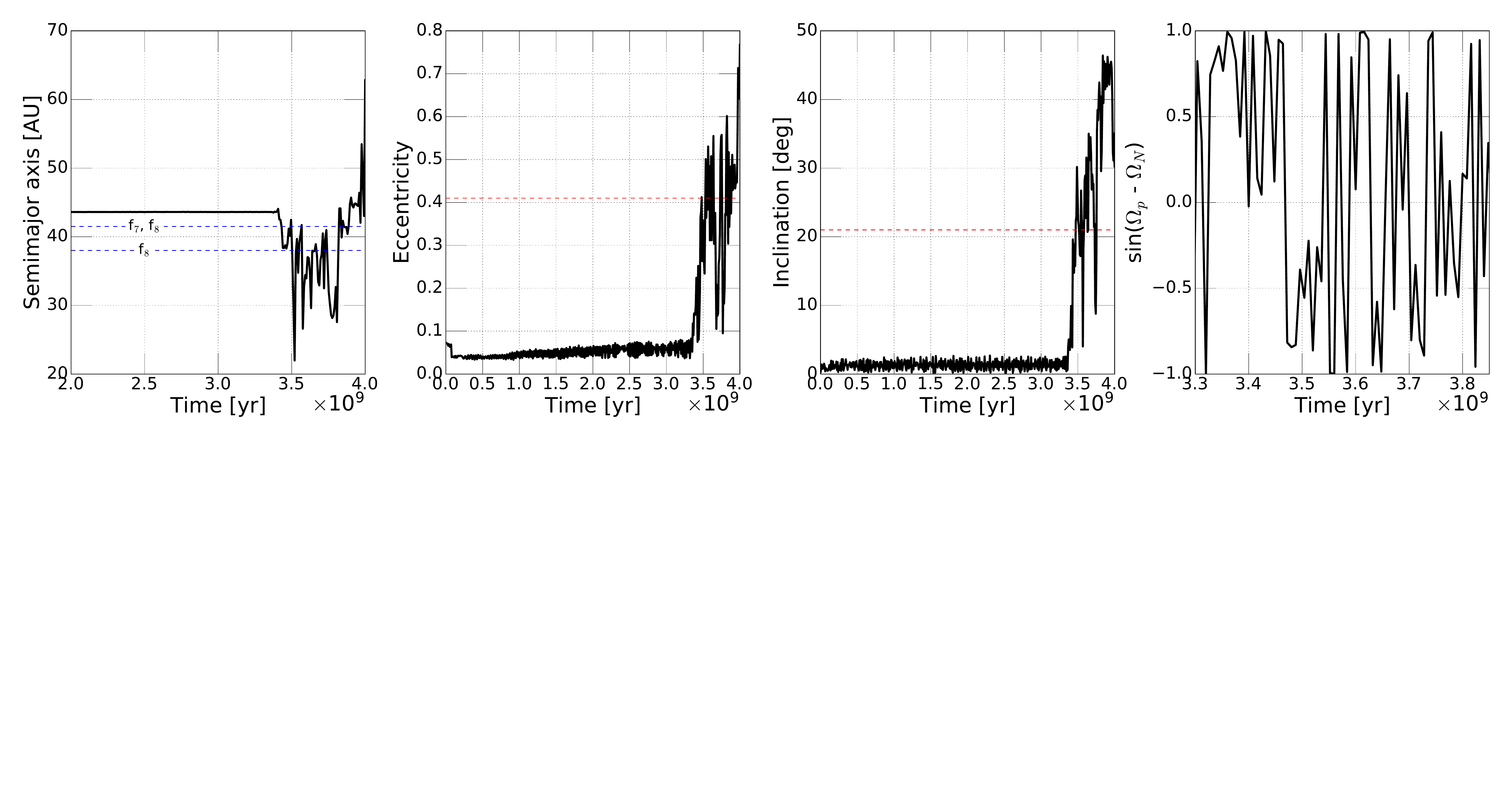}}

    \caption{Time evolution of typical VR particles in our simulations. Top 3 panels are for particles that end up with high eccentricity but low inclination (Q1). Panels 4 and 5 from the top are for cases that end up with high inclination, but low eccentricity (Q3). The bottom panel is for a case that end up with high eccentricity and inclination (Q2). }
    \label{fig:lowIncHighe}
\end{figure*}

First we examine the VR population with e$>$0.42 and inc$<$21 deg (Q1). High eccentricity TNOs usually result from either a scattering event by Neptune or capture in a MMR, thus evolving directly from Q4 to Q1. Manually inspecting these simulated objects reveals that they usually originate as objects captured during Neptune's migration in second and third order MMRs, mainly the 5:2, 5:3, 7:4, and 3:1 resonances. These are then long term stable inside or near the MMRs, sometimes up to a 10$^9$ years, before eventually undergoing a scattering event that put them on their current orbit. The delayed scattering is probably due to the slow and stochastic diffusion of particles inside the MMRs, slowly pushing them out into Neptune crossing orbits. Three representative examples are shown in Fig. \ref{fig:lowIncHighe}. The first case shows a particle originating around 42.5 AU, then getting captured by a migrating Neptune in the 5:2 MMR as evident through its smooth semimajor axis and eccentricity increase, and the libration of 5$\lambda_p$ -2$\lambda_N$ -3$\varpi_p$ resonant angle, where $\lambda$ and $\varpi$ are respectively the mean longitude and longitude of perihelion. Once Neptune's migration has ended, the particle remains captured for $\sim$ 0.5 Gyr, before getting scattered by Neptune into an orbit with a $\sim$ 95 AU and e $\sim$ 0.6. The inclination of this object is never above $\sim$ 15 deg. 
In the second example we show a case where the object is captured in the 7:4 MMR for 2 Gyr, before exiting the resonance, encountering Neptune, and getting scattered to 55 AU , e $\sim$ 0.43, and an inclination of 10 deg. In the final example, the objects is captured into the 3:1 MMR, and remains stable in the resonance for the lifetime of the solar system with e $\sim$ 0.41 and inc $\sim$ 3 deg. It is worth noting however that the 3$\lambda_p$ -$\lambda_N$ -2$\varpi_p$ angle librates only for the first 3$\times 10^8$ yr, before starting to circulate. This indicates that the particle might be on a stable orbit near, but not necessarily strictly inside, the MMR. 

Next we examine the population of VR objects with e$<$0.42 and inc$>$21 deg (Q3). The fundamental mechanism behind this population is Kozai-Lidov oscillations inside MMRs \citep{kozai1,kozai2,kozai3,kozai5,kozai6}, usually the same second and third order resonances seen above. In the first example, we show a particle getting swept by Neptune's 5:2 MMR during its migration, increasing its eccentricity at constant low inclination, before the Kozai-Lidov resonance is triggered as reflected by the libration of the argument of the pericenter around the two stable modes at 90$^o$ and 270$^o$ \citep{kozai4}. The particle's eccentricity and inclination then start to oscillate inversely, periodically increasing the inclination while decreasing the eccentricity, leading eventually to the particle escaping the resonance during a high inclination phase. This mechanism was initially proposed by \cite{gomes2003,gomes2005}. The particle finally ends up with e $\sim$ 0.15 and inc $\sim$ 23 deg. The second example we show follows the same steps, except the particle is caught in the 3:1 MMR, and has higher eccentricity when the Kozai-Lidov oscillations are triggered. This leads finally to e $\sim$ 0.15 and inc $\sim$ 28 deg. We note that in both examples we show the particles never undergo a close encounter, and thus their semimajor axis evolution is smooth, with the particles finishing the simulation on near-MMR orbits. While this behavior is common for the particles in this regime, it is not exclusive. However, if indeed kozai-Lidov oscillations without scattering create this population, then we predict that the binary fraction of these objects should also be higher than for Q1/Q2 where scattering off Neptune is more common.

Some of the particles in our simulation do undergo scattering events, and sometimes even MMR hopping. However the Kozai-Lidov oscillations are present in the majority of cases. We also note that the majority of particles in this category have inclinations relatively close to the 21 deg limit, usually less than 28 deg. 

We finally examine the population of VR objects with e$>$0.42 and inc$>$21 deg (Q2) illustrated through our final example in Fig. \ref{fig:lowIncHighe}. Multiple possible mechanisms contribute to this population. The simplest and most common, shown in our first example, starts with an inward scattering event that both increases the eccentricity of the particle, and allows it to cross the f$_7$ and f$_8$ secular resonances between 38 and 42 AU (reflected through the libration of the longitude of the ascending nodes  $\Omega_p$ - $\Omega_N$), significantly increasing its inclination. The particle is finally scattered again, but outwards to its final orbit. Multiple variations of this mechanism exist, including scattering from inside a MMR after slow diffusion, and Kozai-Lidov oscillations bumping up the particle's inclination moderately before the scattering event and the secular resonance crossing. Since our color transition line is at 42 AU, and all major secular resonances capable of inclination excitation are interior to it, then only particles scattered inward can cross these resonances. The scattering naturally increases the particle's eccentricity, leading to the strong correlation between very high inclination (inc $>$ 28 deg) and high eccentricity VR particles.  

\begin{figure}
    \centering
    {\includegraphics[scale=0.50]{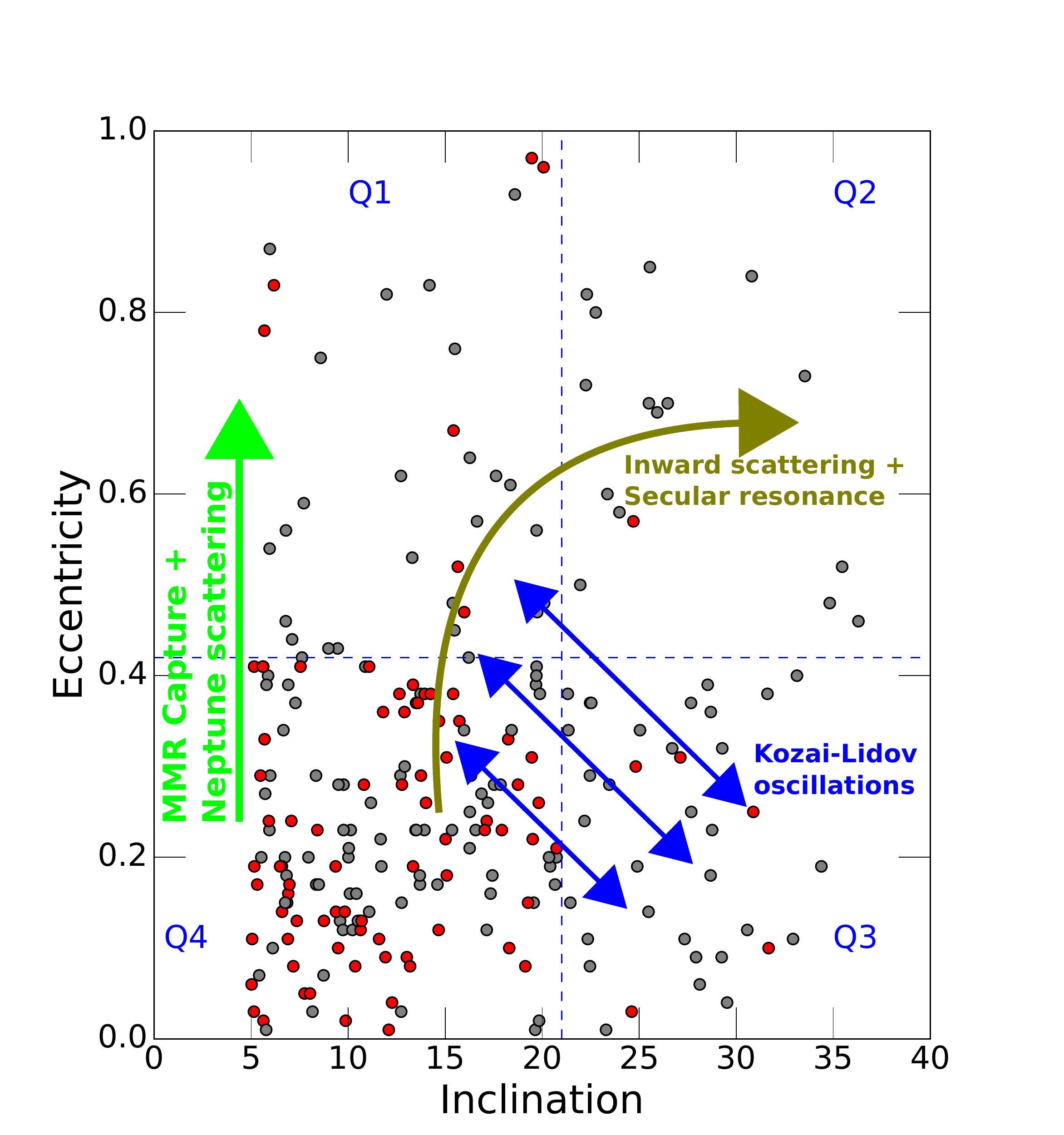}} 

    \caption{A summary of the mechanisms responsible for red objects on high eccentricity or inclination orbits. Quadrant Q1 objects with high eccentricity but low inclination are initially captured in second and third order MMRs, before escaping the resonance and getting scattered outward by Neptune. Quadrant Q3 red objects with high inclination but low eccentricity start as either Q4 or Q1 objects, before undergoing Kozai-Lidov oscillations (usually inside the MMRs) that lift their inclinations while decreasing their eccentricities. Finally, Q2 objects with both high eccentricity and inclination also start by getting captured in a MMR, but then undergo an inward scattering by Neptune, allowing them to cross the f7 and f8 secular resonances, raising their inclination. They are finally rescattered outwards to their final orbits. }
    \label{fig:mechanisms}
\end{figure}

We summarize the three mechanisms responsible for red high eccentricity or inclinations objects in Fig. \ref{fig:mechanisms}.

\subsection{Effects of granular migration}
Our nominal model in the previous section assumed smooth planetary migration. In this section we investigate the effects of granular migration for Neptune. {In the models of \cite{nesv2016,kaib2016}, granular migration significantly increases the number of particles that fall out of mean motion resonances. We add granularity to our nominal model using a simple prescription based on that of  \cite{kaib2016}, who used a parametric fit to the N-body results of \cite{nesv2016}.}

We create a distribution of roughly 20,000 positional jumps amplitudes $\delta_x$ [AU] by sampling the following Gaussian:
\begin{equation}
f(\Delta a)=\frac{1}{\sqrt{2 \sigma^{2} \pi}} \exp -\frac{(\Delta a-\mu)^{2}}{2 \sigma^{2}}
\end{equation}  
where $\mu=\pm 3.75 \times 10^{-4} \mathrm{au} \text { and } \sigma=1.8 \times 10^{-3} \mathrm{au}$. Notice that the Gaussian mean is a factor $\sim$ 6 less than the value used  by \cite{kaib2016}, since we find that higher values lead to very high dynamical noise and chaotic evolution for Neptune, significantly more than in \cite{kaib2016}. This is probably due to the different integrators used. We moreover set a maximum amplitude of $\pm$ 7.5 $\times 10^{-4}$ AU. We generate the jumps encounter times similarly to \cite{kaib2016} by
assuming that they are uniformly distributed until t=$\tau_m$/10, and then they fall off as $t^{-1.15}$ until t=3$\tau_m$. 
We finally apply the jumps to Neptune's cartesian coordinate component $x$ as:
\begin{equation}
\texttt{Neptune.x} = \texttt{Neptune.x} + \delta_x
\end{equation}

\begin{figure*}
    \centering
    \subfigure{\includegraphics[scale=0.30]{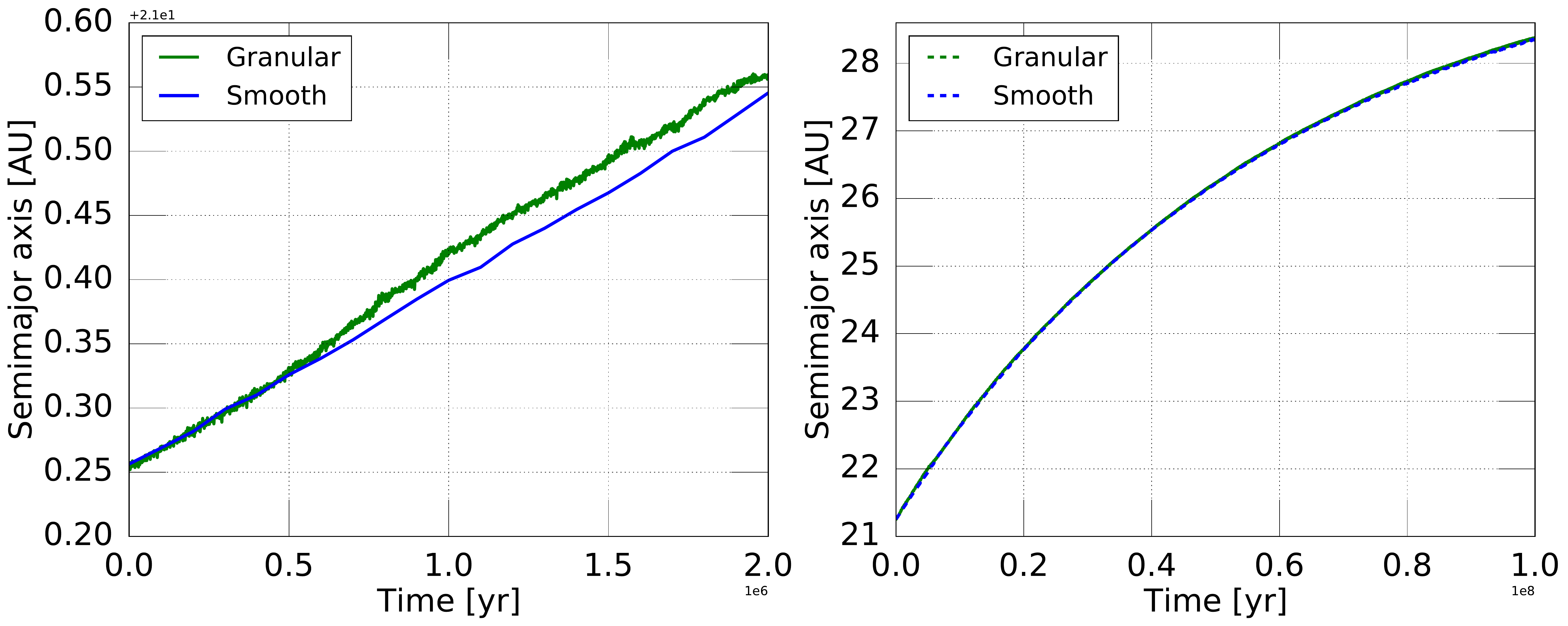}} 
    \subfigure{\includegraphics[scale=0.30]{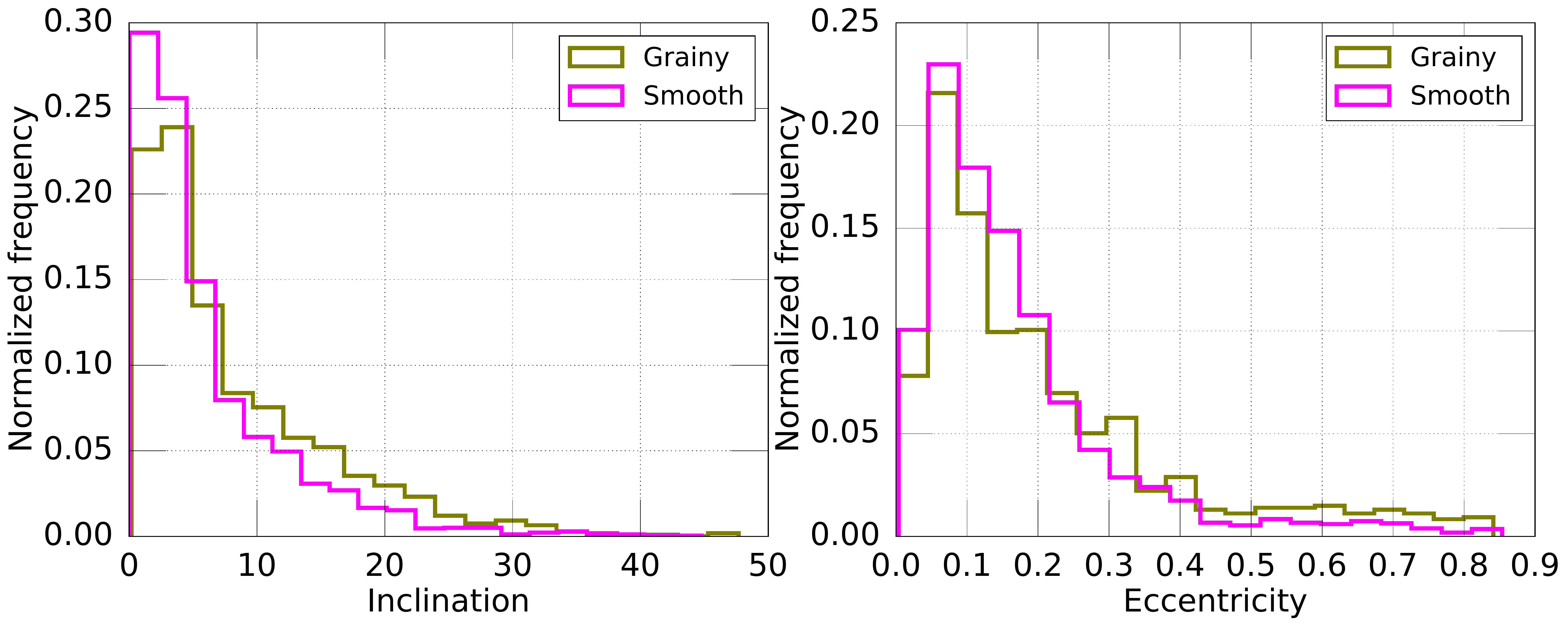}} 
    \subfigure{\includegraphics[scale=0.30]{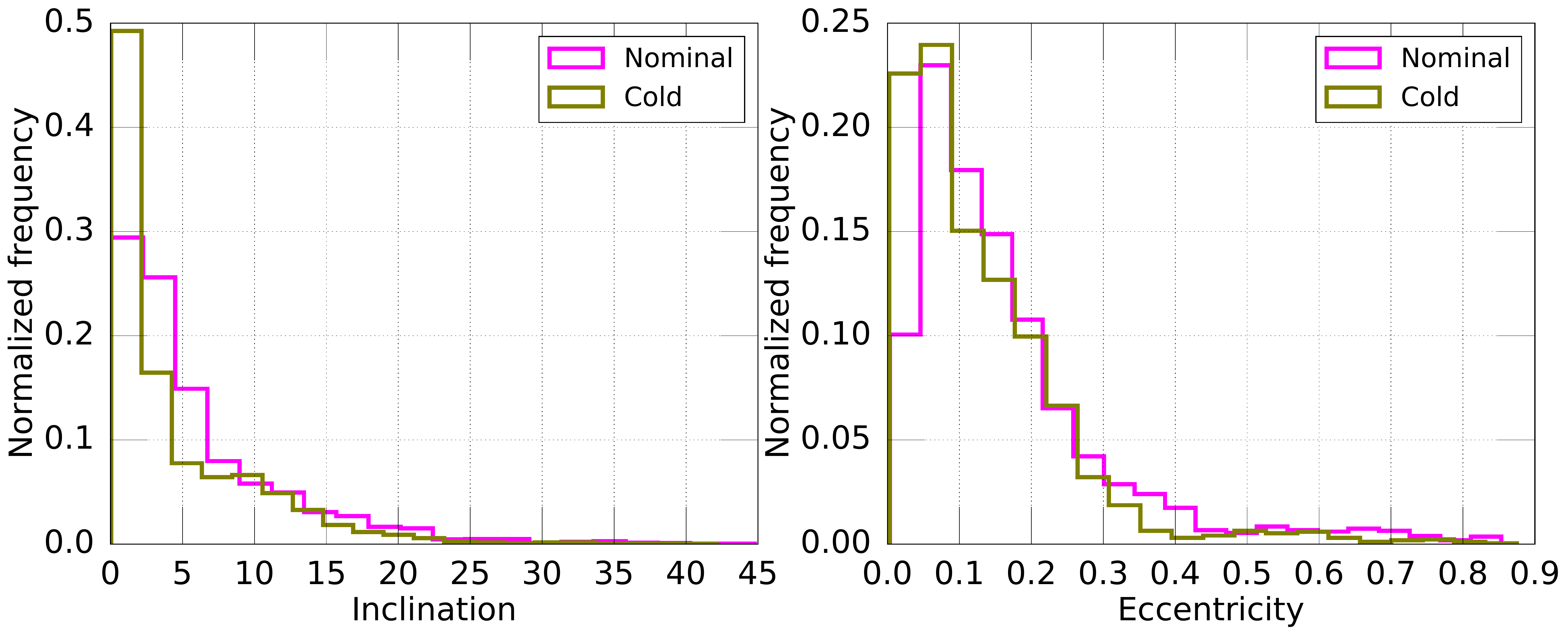}} 

    \caption{Top panels: Neptune's semimajor axis evolution in our grainy model, compared to the smooth model. Middle panels: comparisons of the eccentricity and inclination frequency histograms of the two models. Bottom panels:  Comparisons of the eccentricity and inclination frequency histograms between our nominal model, and an initially dynamically colder case. }
    \label{fig:granular}
\end{figure*}

This contrasts with \cite{kaib2016} who applied the jumps to the semimajor axis directly. The main reason for using the cartesian coordinates is because, in \texttt{Rebound}, orbital elements such as the semimajor axis are protected attributes for the class member, and hence we cannot change them directly during a simulation from within \texttt{Python} API. An alternative would be to create a \texttt{Reboundx} \textit{operator} (instead of a force) that can directly act on the semimajor axis. The issue with this approach is that the optimal integrator for our setup (\texttt{Mercurius}) is incompatible with \texttt{Reboundx} operators. We hence choose to apply our jumps to \texttt{Neptune.x} directly, while turning on the \texttt{whfast.safe$\_$mode} flag, turning off the symplectic correctors, and making sure that the particles' positions and velocities are synchronized. Comparison of our smooth and granular models is shown in the top panels Fig. \ref{fig:granular}. While Neptune's two semimajor axis evolution curves are practically identical on large (AU) scale, the grainy migration curve is significantly more jittery at the 0.01 AU scale. We note that due to these jumps,  the giant planets evolution here is not bit by bit identical to the smooth case. While we start the system with the exact same initial conditions, and the final semimajor axis, inclinations, and Uranus' eccentricity are very close to the smooth migration case, Neptune's final eccentricity is a factor $\sim 2$ higher. 

In the bottom panels of Fig. \ref{fig:granular} we compare the final test particles eccentricity and inclination distributions for both models. We find that grainy migration, in the limits of our implementation, leads to a factor $\sim 1.5$ more high eccentricity (e$>0.22$) and inclination (inc $> 8$) particles than smooth migration, while equally decreasing the fraction of dynamically colder particles. Inspecting the LR to VR particles number ratio, assuming our nominal color transition line at 42 AU, we find 0.29 and 1.91 respectively for inside and outside of Q4. These numbers are practically identical to those obtained in the smooth migration case (0.27 and 2.0), indicating that the extra dynamically hot particles obtained here preserve the color ratios.

\begin{figure*}
    \centering
    \subfigure{\includegraphics[scale=0.30]{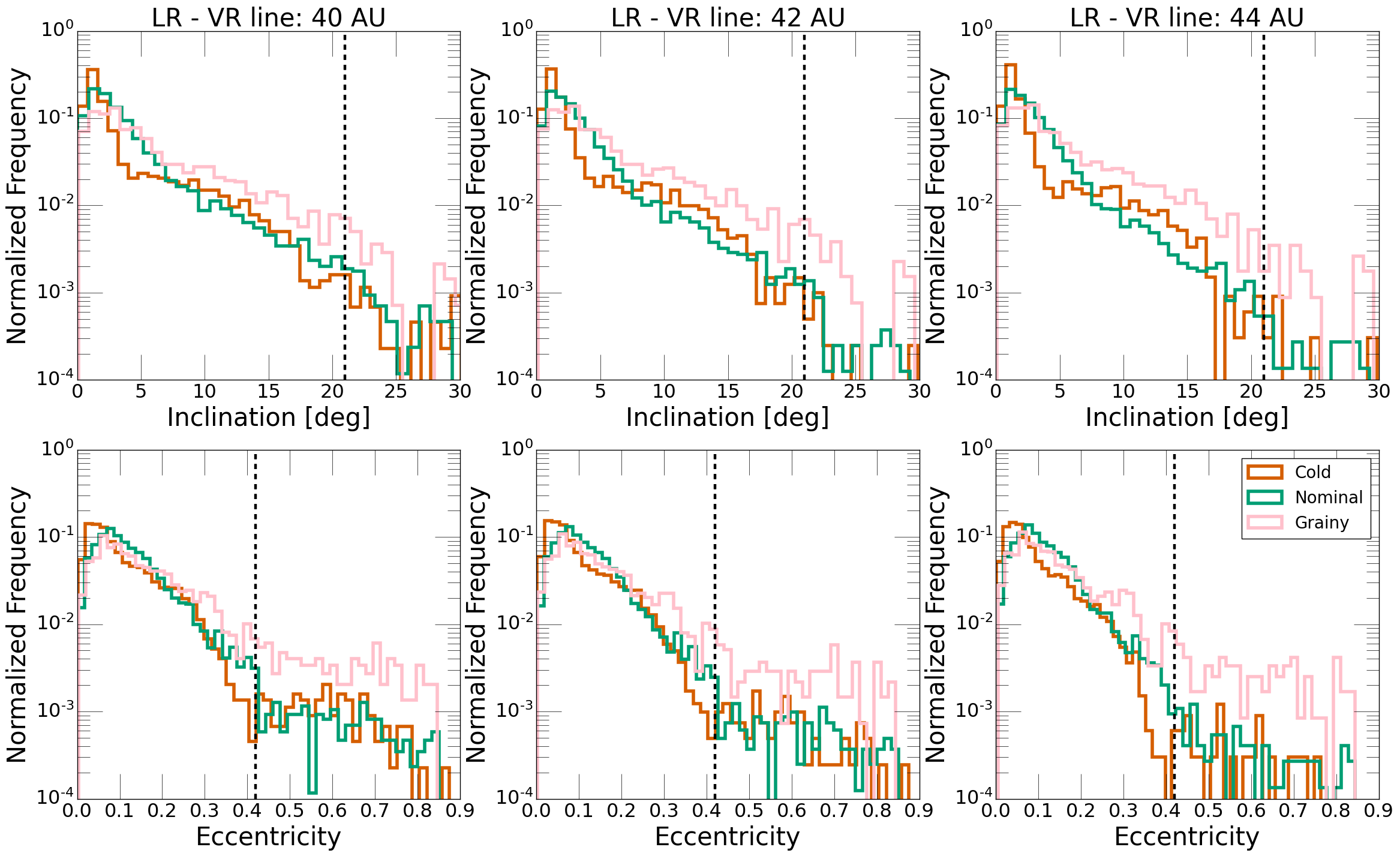}} 

    \caption{Top panels: Final inclinations normalized frequency histograms of VR particles in our nominal, cold, and grainy models, for different LR-to-red transition locations. Here we exclude VR particles with eccentricities higher than 0.42. Bottom panels: Final eccentricities normalized frequency histograms of VR particles in our nominal, cold, and grainy models, for different LR-to-red transition locations. Here we exclude VR particles with inclinations higher than 21 deg. }
    \label{fig:granular2}
\end{figure*}

In Fig. \ref{fig:granular2} we plot the VR particles eccentricity and inclination normalized frequency distributions for our granular model, compared to the nominal case, considering three different color transition lines. We notice that grainy migration leads to more VR high inclination or eccentricity particles in almost all cases, and while it does seem to slightly push the high inclination cliff outwards into the cold classical belt, it does not affect the location of the eccentricity distribution plateau.

We hence conclude that the granularity of Neptune's migration has little effects on our results, considering the sparsity of the observational data. We however emphasize that: 1- higher levels of granularity might change our conclusion, and 2- granular migrations have important effects for the population of individual resonances, and might hence be relevant for future more complete color observations.

\subsection{Effects of the disk's initial eccentricity and inclination}
Our choice to initiate the simulation with a moderately stirred-up proto-Kuiper belt is motivated by many studies showing possible primordial TNOs disk excitation due to the earlier dynamical history of the 4 gas giants \citep{chiang,tsiganis,hahn2005}. These studies moreover showed that, as initially discussed by \cite{dermott}, dynamically hot particles have a higher probability of being captured into Neptune's higher order MMRs, such as 3:1 and 5:2, during its migration.

To investigate the effects of starting with a dynamically colder disk, we ran an otherwise identical set of simulations where we sample the particles disk from a Rayleigh eccentricity distribution with a scale of 0.04, and a Rayleigh inclination distribution with a scale of 0.02 rad, both a factor 2 below our nominal model. Comparisons of the final eccentricity and inclination distributions of VR particles exclusively are also plotted in Fig. \ref{fig:granular}. This show significantly more particles with e $<$ 0.05 and inc $<$ 2.5 deg for the cold case, while the nominal case has more particles in the moderately excited range (e $\sim$ 0.12 and inc $\sim$ 5 deg). The cold case also seems to have less particles on very hot orbits, but the statistic is too low for a definitive conclusion in this regard. We finally compare the LR to VR ratios between the two cases, and find values of 3.77 and 0.27 for outside and inside Q4, respectively. The cold disk case therefore leads to the same LR-to-red ratio inside Q4 as the nominal case, but for more VR particles outside of it.

In Fig. \ref{fig:granular2} we plot the VR particles eccentricity and inclination normalized frequency distributions for our cold model, compared to the grainy and nominal cases, considering three different color transition lines. We notice that the cold model have the same inclination cliff and eccentricity plateau location as the hotter model, and therefore is consistent with the same color transition location derived from our nominal case.

\subsection{Effects of the disk's initial surface density profile}
\begin{figure*}
    \centering
    \subfigure{\includegraphics[scale=0.22]{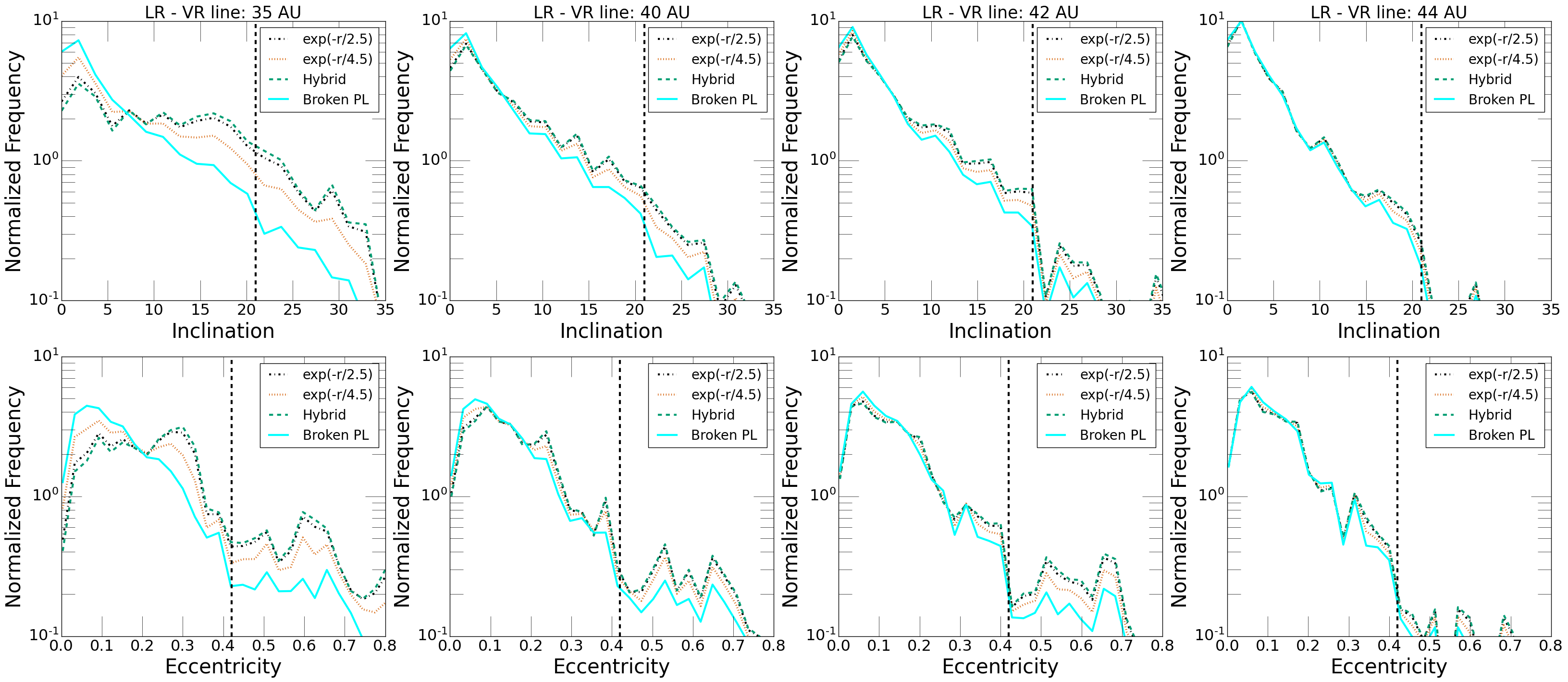}} 
    \caption{{The eccentricity and inclination distributions of VR objects for different disk initial surface density profiles and color transition locations. Plotted are 4 different profiles: 2 exponential profiles with different e-folding distances, a truncated power laws case, and a hybrid case.}}
    \label{fig:diskdensity}
\end{figure*}

{ The final parameter we explore is the initial surface density profile of the planetesimals disks. In our nominal model above we used $\Sigma \propto r^{-2}$ as it reproduces reasonably well many aspects of the dynamical and color structures of the Kuiper belt. This profile however suffers from two possible issues: 
\begin{itemize}
    \item it might not be consistent with the mass of the cold classical belt, as it implies that the ratio between the mass available at 30 and 45 AU is only 1.5. Assuming a mass of $3\times10^{-4} M_{\oplus}$ for the cold classicals at 45 AU \citep{fraser2014} (or $\sim 10^{-4} M_{\oplus}$ / AU), this imply a mass at 30 AU $\sim 10^{-4} M_{\oplus}$ / AU. This is orders of magnitude less than the $\sim 1 M_{\oplus}$ / AU at 30 AU needed for Neptune's migration \citep{nesvmorb}. If indeed $\Sigma \propto r^{-2}$, this would mean that either the currently observed mass of cold classicals does not constrain the mass initially available around 45 AU, or that cold classicals were depleted by 4 orders of magnitude. Note that an alternative possibility is a heavy disk that ends at 42 AU, with a cold belt implemented later via the mechanism of \cite{gomes2020}.
    \item our conclusion that color transition lines below 38 AU produces too many VR objects to be consistent with observations might be surface density profile dependent. Other disk profiles, with less mass available around 30-38 AU, can possibly allow us to relax this constraint. 
\end{itemize}
    
We will hence investigate the effects of using different initial surface density profiles. As running full simulations for each case is numerical prohibitive, we will follow \cite{nesv2020} in using a weighting scheme with our nominal model to mimic a different initial disk profile. We hence assign each object a weight $w(r)$ that follows the selected surface density profile. The weights for the nominal (1/$r^2$) model particles are $w_{nom}(r)=1$ everywhere, and hence each particle is only counted once in this case. For the weighted cases, each particle is counted $C_w \times w_{nom}(r) \times r^2 \ \Sigma_{new}(r) $ times, where $C_w$ is a normalization constant. For surface density profiles steeper than our nominal case, $w(r)$ will be $>$ 1 at small semimajor axis, and $<$ 1 in the outer disk.
Note that this method in principle is inapplicable to particles starting inside 30 AU, as their history is highly coupled to Neptune's migration. Since our LR-VR transition line is never inside 35 AU, this method should be reasonably accurate when studying the VR objects. 

We also follows \cite{nesv2020} in exploring three different surface density profiles: $\Sigma(r) = \frac{1}{r} \exp{[- (r - 22)/2.5]}$, a truncated power law with $\Sigma \propto 1 / r^{\gamma}$ with $\gamma$=1.4 inside 30 AU, and $\gamma$=1 outside of it, with a scaling factor of 1000 at 30 AU, and finally a hybrid case with a power law inside 30 AU, and exponential decrease outside. We also add an additional case with $\Sigma(r) = \frac{1}{r} \exp{[- (r - 22)/4.5]}$, that while still underpredict the cold classicals mass, we find it to be an interesting case that leads to smaller ratios than the $\Sigma(r) \propto \exp{[-r/2.5]}$ case.

We first start by investigating the possible presence and location of eccentricity and inclination cutoffs for VR objects, as this is the most robust method to compare the models to the observations. In Fig. \ref{fig:diskdensity} we plot the final eccentricity and inclination normalized frequency distributions of VR objects for the 4 weighted initial profiles, and for 4 different color transition line locations. We first notice that the 4 curves are partially correlated, which is a consequence of using a weighting scheme. Looking closer, we notice that while none of the curves show an inclination cutoff at 21 deg for transitions at 35 or 40 AU, all 4 show a cutoff to a certain extent for a transition at 42 AU. For the eccentricity, we notice that all cases also show a plateau for transitions at 40 and 42 AU. All in all, while the 4 profiles do lead to seemingly different distributions, none of them necessitate a color transition location different than our nominal model (40 to 42 AU).

As our previous analysis showed no significant effects for using different initial surface density profiles on the VR particles e-inc cutoff, We now follow up our by calculating the VR-to-LR ratio inside and outside of e = 0.42 and inc = 21 deg for the different profiles, and inspecting VR objects inside the 3:2 MMR. As discussed for the nominal case, we reemphasise that these numbers should be interpreted very loosely, as all of the observed ratios are biased and might not be reflective of the intrinsic values even after correcting for the different albedos and other effects. They however do offer valuable informations when comparing the \textit{different models together}. Note that to obtain the reported ratios, the weighting was done on the biased population obtained through the Survey Simulator, as for our nominal model. The final ratios for the different profiles are summarized in table \ref{tabres}. We report the LR-to-VR ratios inside and outside of ecc = 0.42 and inc = 21 deg.  Our results show that, overall, the two exponential profiles with color transition at 35 AU leads to closer ratios compared to observations than the nominal case, with the added advantage over a 40-42 AU color transition of populating the 3:2 MMR with VR particles. The $exp(-r/2.5)$ case however also lead to a factor 5-10 less VR particles in this resonance than the nominal and $exp(-r/4.5)$ cases, for the same transition location. If the ratio observed by M2019 in the 3:2 is any reflective of the intrinsic value the $exp(-r/2.5)$ case would not be a good solution, even excluding that it does not fit the eccentricity-inclination cutoff. The acceptable color ratios for these profiles with transitions at 35 AU, combined with our results in Fig. \ref{fig:diskdensity} showing that such transition does not reproduce sharp e-inc cutoffs in the VR population, imply that the exponential profiles are leading to a smooth decline in the VR population as a function of the eccentricity and inclination. While the exponential profiles might be consistent with the observed color ratios in SDOs and plutinos, they are not consistent with the sharp constrains on VR objects to inside e=0.42 and inc=21 deg.

Inspecting the truncated power law profile case, we find that it always leads to ratios inside and outside the ecc-inc cutoff that are much closer than the other cases. Finally, the hybrid case always leads to ratios much higher than the other cases, sometimes by an order of magnitude. 

From all of these 3 lines of evidences, we conclude that using different initial surface density profiles does not seem to eliminate the tension between the rarity of VR SDOs, the abundance of VR objects in the 3:2 MMR, and possibly the color ratios contrast inside and outside of ecc = 0.42 and inc = 21 deg. Assuming that full simulations with different initial surface density profiles behave similarly to our weighted profiles, and that the unbiased color ratios are inline with our dataset's values, this hints that a more complex migration history for Neptune than assumed here might be needed to solve this contradiction. }

\begin{table}[]
\begin{centering}

\begin{tabular}{llllll}
\hline
\multicolumn{1}{|l}{{Profile}} &                                & \multicolumn{4}{c|}{{LR-VR   transition [AU]}} \\ \hline
 &  & \multicolumn{1}{c}{{30}} & \multicolumn{1}{c}{{35}} & \multicolumn{1}{c}{{40}} & \multicolumn{1}{c}{{42}} \\ \cline{3-6} 
                                      &                                &            &            &           &            \\ \cline{1-1}
& &      {LR/VR} & {LR/VR}  &     {LR/VR}     & {LR/VR} \\     
{1/r$^2$ full sim}                & {Inside 0.42 - 21 deg}  &  0.06       &  0.065      &  0.18      &  0.26       \\
{}                             & {Outside 0.42 - 21 deg} & 0.08       & 0.083      & 0.62      & 2.0          \\
{}                             & {}                      &            &            &           &            \\ \cline{1-1}
{exp(-r/4.5) weighted}         & {Inside 0.42 - 21 deg}  & 0.64       & 0.7        & 1.72      & 1.56       \\
{}                             & {Outside 0.42 - 21 deg} & 0.225      & 1.4        & 2.98      & 10.59      \\
{}                             & {}                      &            &            &           &            \\ \cline{1-1}
{exp(-r/2.5) weighted}         & {Inside 0.42 - 21 deg}  & 4.1       & 5.04       & 18.3     & 15.6     \\
{}                             & {Outside 0.42 - 21 deg} & 0.6       & 12.5      & 21.1     & 73.7     \\
{}                             & {}                      &            &            &           &            \\ \cline{1-1}
{truncated PL weighted}           & {Inside 0.42 - 21 deg}  & 2.3       & 2.05       & 3.3      & 1.5       \\
{}                             & {Outside 0.42 - 21 deg} & 2       & 3.6       & 6.1      & 1.7       \\
{}                             & {}                      &            &            &           &            \\ \cline{1-1}
{hybrid weighted}              & {Inside 0.42 - 21 deg}  & 5.05       & 8.2        & 40        & 38         \\
                                      & {Outside 0.42 - 21 deg} & 0.73       & 24.9      & 45        & 157        \\
        \hline
\end{tabular}
\caption{The LR-to-VR objects ratio inside and outside of the ecc = 0.42 and inc = 21 deg constrain for different initial surface density profiles and color transition line locations. }
\label{tabres}

\end{centering}
\end{table}



\section{Summary and discussions}
\label{conclusion}
\subsection{Summary}
\begin{enumerate}
    \item By reanalyzing the dataset of \cite{marsset} that includes data from Col-OSSOS and other surveys, we show that VR TNOs are strongly limited to eccentricities less than $\sim$ 0.42. This result, taken together with the color-inclination correlation of \cite{marsset}, implies that VR TNOs are constrained in e-inc space to inside e=0.42, and inc = 21 deg. 
    \item Inspecting the individual populations in the dataset, we show that the color-eccentricity and color-inclination correlations are created by different bodies, with the dominant LR colors of SDOs and high order MMRs creating the first, and multiple different populations contributing to the second. We however find that the 3:2 MMR contains both LR and VR objects in significant amounts.
    \item Using N-body simulations accounting for Uranus and Neptune's migration, we show that the data is best explained through a primordial color transition line around 38-42 AU in the protoplanetary disk, with LR and VR objects forming inside and outside of it, respectively. This is consistent with the results of  \cite{nesv2020} who hypothesized a roughly similar color transition line to explain the results of \cite{marsset}. {Our nominal model leads to LR-to-red objects ratio of 0.27 and 2.0 inside and outside of e=0.42 and inc = 21 deg, respectively. This is a factor $\sim 3$ within the observational values where LR-to-VR ratios are $\sim$ 1.04 and 5.84.}
    \item  We however find a possible tension between the rarity of the VR objects in the scattered disk, and their abundance in the 3:2 MMR, as a color transition line inside 38 AU is needed to explain the latter, but this also leads to a large number of VR objects in the scattering disk. {This tension is not alleviated when using different disk initial surface density profiles, although this was investigated using weighted profiles instead of full simulations.  }
    \item We additionally find that adding modest graininess to the migration does not change the LR-to-VR ratios of our nominal model, as both VR and LR particles are equally affected by this graininess.
    \item We finally investigate the origins of the VR objects outside of the e=0.42 and inc = 21 deg quadrant, and find three distinct mechanisms behind these objects in the three different parts of phase space. $\bullet$ VR objects with e $>$ 0.42 and inc $<$ 21 deg usually start as particles trapped in second and third order MMRs for 10$^8$-10$^9$ yr, before exiting the resonance, and getting scattered into higher eccentricity orbits by Neptune. $\bullet$ VR objects with e $<$ 0.42 and inc $>$ 21 deg also start as particles trapped in the same MMRs, but then undergo Kozai-Lidov oscillations inside the resonances, raising their inclination while lowering their eccentricity. This allows them to escape the MMR and remain on the high inclination, low eccentricty orbits. $\bullet$ VR objects with e $>$ 0.42 and inc $>$ 21 deg yet again start as particles trapped in MMRs, including the same second and third order resonances, but also the 2:1 MMR. These objects eventually escape the resonance, and get scattered \textit{inwards} by Neptune, thus crossing the f7, f8, g7, and g8 secular resonances. The scattering itself raises their eccentricity,  while the secular resonances crossing increases their inclination dramatically, putting them on high eccentricty and inclination orbits.
\end{enumerate}

\subsection{Discussions}
\begin{enumerate}
    \item While we hope that our results will contribute to understanding the early evolution of the solar system, this approach can benefit greatly from both more data and simulations. For example, the limited scope and size of the M2019 dataset did not allow us to inspect the precise LR-to-red ratios in the 4 individual e-inc space quadrants (Fig. \ref{fig:ecc3}), or in individual MMRs.   
    \item {The numerical model we uses in this work can moreover be improved. While we did check smooth and granular migration, in addition to cold and hotter initial TNOs disks, we tried only a single migration timescale value. We moreover did not inspect migration schemes with ``jumping Neptune'' \citep{nesv2015a}, or short range transport \citep{gomes2020}. We moreover fully simulated a single particles number density distribution, while other distributions (exponential, truncated power law) were mimicked via a weighting scheme. Exploring these models with full simulations might shed more light the tension between the abundance of VR objects in the 3:2 MMR, and their absence in the scattered disk.}
    \item In our granular model, we used significantly lower jump mean amplitude than \cite{kaib2016}, as higher values led to a chaotic evolution for Neptune. While this can simply be due to the different codes (and algorithms) used by the two groups, fully understanding the origins of this discrepancy is of interest for future works. 
    \item The disk physical-chemistry interpretation of the color transition line at $\sim$ 38-42 AU is also an open question. The simplest such transition imaginable would be the outermost snowlines (CO or N$_2$). It is however hard to push any of these this far out without a very shallow temperature profile, where CO snowlines are usually observed to be inside 30 AU \citep{2016ApJ...823...91S}, although larger values are also possible \citep{qi}. Other possibilities are the NH$_3$ or H$_2$S retention lines as proposed by \citep{brown}. Further constraining the location of the color transition line using TNOs color observations can hence shed light on the physical-chemical structure of the protoplanetary disk. 
\end{enumerate}

\section*{Acknowledgements}
We thank the referee for their useful comments that helped us improve our manuscript significantly. We thank N. Kaib, A. Morbidelli, K. Öberg, C. Petrovich, and K. Volk for many useful discussions on the dynamics and chemistry of the Kuiper Belt. We also thank H. Rein  and D. Tamayo for answering our \texttt{Rebound} related questions. M.A.-D. was supported through a Trottier postdoctoral fellowship at the University of Montreal, followed by a CAP3 fellowship at NYU Abu Dhabi. The computations were performed on the Sunnyvale cluster at the Canadian Institute for Theoretical Astrophysics (CITA), and Compute Canada's Béluga and Graham clusters.






\end{document}